**Title:**

High-fidelity six-dimensional optical data storage

**Authors:**

Jichao Gao[1*], Tong Jin[1*], Chenqi Yi[1*], Yihui Fu[1], Zhi Yan[1], Xiya Xie[1], Rokas Drevinskas[2], Zhexiang Xiao[1], Lei Wang[3], Guangda Niu[1✉], Jingyu Zhang[1,2✉], Hong-Bo Sun[4✉]

**Affiliations:**

1. Wuhan National Laboratory for Optoelectronics, Huazhong University of Science and Technology, Wuhan 430074, China
2. Yiotta Tech. Co. Ltd., Wuhan 430070, China
3. State Key Laboratory on Integrated Optoelectronics, Jilin University, Changchun 130012, China
4. Tsinghua University, Beijing 100084, China

*These authors contributed equally to this work.

✉e-mail: guangda_niu@hust.edu.cn; jy_z@hust.edu.cn; hbsun@tsinghua.edu.cn

**Abstract:**

**Multi-dimensional optical data storage, wherein each voxel encodes multiple bits within a three-dimensional isotropic medium, presents a revolutionary approach to dense data storage by exploiting intensity, polarization, and wavelength to control distinct optical responses. However, increasing dimensionality exacerbates crosstalk and compromises recording fidelity. Here, we introduce a six-**

**dimensional (6D) multiplexing approach using ultrafast interactions of polarized light with a three-dimensional medium. By manipulating three light parameters within a Poincaré-like semi-sphere, we induce three material responses in laser-induced nanostructures—fluorescence, phase retardance, and slow axis orientation—each corresponding to a unique physical quantity. This one-to-one correspondence between light parameters and material responses achieves high fidelity with 99.45% raw accuracy. Furthermore, independently controlled 6D multiplexing maximizes effective user capacity to 0.38 petabits per disc. The versatility of laser-induced nanostructures across physical dimensions enables applications from high-capacity data storage to microfluidics, optical elements, integrated photonics, micromechanics, and quantum technologies.**

**Main text:**

Multi-dimensional optical data storage (MDODS) techniques, known for their high capacity, eco-friendliness, and longevity, have attracted substantial interest[1]. The primary challenge in MDODS is achieving high fidelity alongside high-order multiplexing, due to signal interference among distinct physical dimensions. The strategic deployment of multiple light parameters to manipulate an equivalent number of material responses and corresponding physical quantities establishes a robust correspondence between the dimensions of light and material (Figure 1a)[2–6]. This direct correlation leads to high fidelity, thereby maximizing effective user capacity and ensuring the system's practicality, in accordance with Shannon's noisy coding theorem[7].

However, constrained by the capabilities of light manipulation methods and recording materials that facilitate interactions in multiple orthogonal physical domains[2,8,9], most technologies beyond four dimensions suffer from a sharp decline in fidelity, leading to numerous issues such as reduced storage density and system complexity. Therefore, achieving high fidelity in optical multiplexing beyond four dimensions remains a substantial challenge that urgently needs to be addressed (Figure 1a).

Plasmonic nanostructures, renowned for their capability to confine light-matter interactions at the nanoscale, are crucial in on-chip photonics[10] and MDODS[8,9]. Although diverse light parameters such as wavelength[8,9], polarization[8,9], and orbital angular momentum[9] are utilized in plasmonic MDODS, their effects are limited to altering the morphology of plasmonic nanoparticles (Figure 1a, inset iii). During readout, all physical properties are inferred from this single material change, exclusively manifested as fluorescence signals. This reliance on a single material response significantly exacerbates signal interference, making high-fidelity readouts a formidable challenge (Methods).

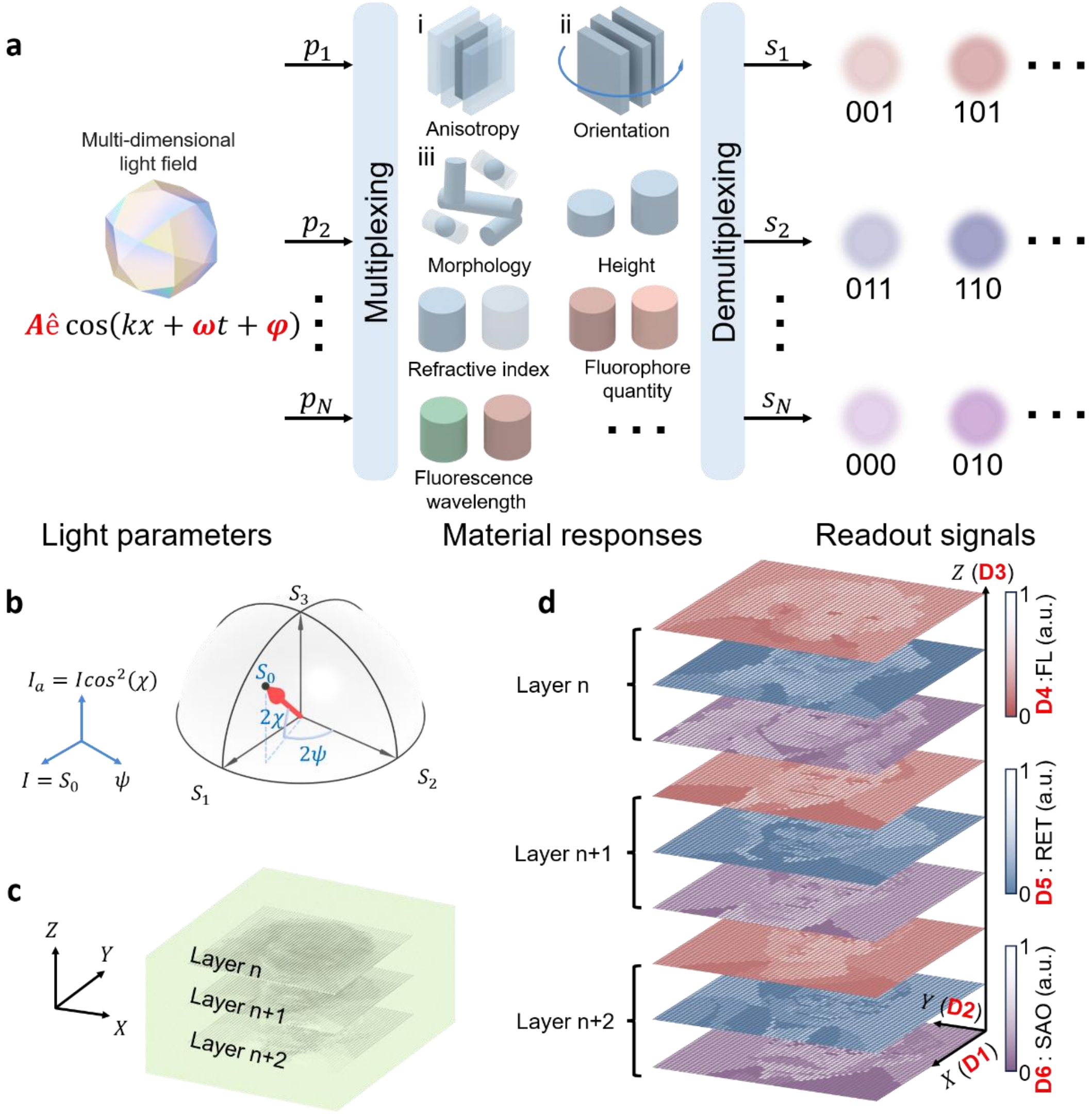

Figure 1 Schematic overview of MDODS data writing, storage, and decryption processes. **a**, Schematic representation of MDODS. Light multiplexed within a multi-dimensional space induces diverse material responses. These material responses are then separately demultiplexed into distinct readout signals. Key light parameters involved are highlighted. Symbols $A,\ \hat{e},\ k,\ \omega,\ \varphi,\ p,\ s$ denote amplitude, polarization, wavenumber, angular frequency, phase, multiplexed light parameter, and readout signal, respectively. **b-d**, Schematics of 6D multiplexing showing (**b**) three manipulated light parameters within a Poincaré-like semi-sphere parameter space; (**c**) data voxels arranged in a 3D material space; and (**d**) Nine images of prominent scientists encoded across the spatial three-dimensional (X-Y-Z; 1$^{st}$-3$^{rd}$ dimensions, D1-D3) multiplexed data layers using fluorescence intensity (FL; 4$^{th}$ dimension, D4), retardance (RET, 5$^{th}$ dimension, D5), and slow axis orientation (SAO; 6$^{th}$ dimension,D6). The $S_0$, $S_1$, $S_2$, $S_3$ represent the Stokes parameters. $S_0$, $\psi$ and $\chi$ correspond to the radius of the semi-sphere, azimuthal angle and ellipticity angle, respectively.

Another approach leverages the form birefringence inherent in self-assembled bulk nanostructures[11]. These nanostructures facilitate permanent five-dimensional optical data storage within glass[2]. By using two light parameters, intensity ($I$) and polarization, to control structural anisotropy and orientation (Figure 1a, inset i and ii), respectively, a robust double one-to-one correspondence is established, achieving nearly 100% fidelity in systematic verifications [2,12,13]. Further extending the number of multiplexed dimensions beyond five presents substantial challenges, and ensuring high fidelity within this increased complexity is even more formidable. Such challenges frequently arise across various MDODS techniques, predominantly due to the one-to-multi correspondence between light parameters and material responses. For example, intensity serves as one of the most important light parameters, yet it simultaneously influences a variety of material responses and their corresponding physical quantities, including RET[14], FL[6], fluorescence spectra[5], refractive index[15] and structural morphology[16]. Such one-to-multiple coupling between light parameters and material responses constrains the expansion of storage dimensionality, as changes in a single laser parameter simultaneously affect multiple material properties, resulting in crosstalk and reduced fidelity.

We have crafted an approach to manipulate the order of nonlinear absorption by engineering the effective bandgap of the medium. By suppressing electron dephasing during multiphoton excitation (MPE), this approach decouples coupled material responses and establishes a triple one-to-one correspondence between light parameters and material responses. Specifically, within this approach, three light parameters are modulated as inputs (Figure 1b), inducing three corresponding material responses that serve as data carriers. The signals retrieved from these material responses then function as outputs. This methodology facilitates the development of a robust, high-accuracy system for data writing, storing, and readout. Within the three-dimensional (3D) space of the recording medium (Figure 1c), we successfully demonstrate 6D multiplexing using FL, RET and SAO with exceptional fidelity (Figure 1d).

The key to devising a high-fidelity, high-dimensional data storage scheme lies in establishing a system that facilitates multiple one-to-one correspondence between light parameters and material responses. This capability enables the mapping of the multi-dimensional space defined by light parameters onto a corresponding material space comprised of orthogonal physical quantities. Such an arrangement ensures that each data channel can be decoded independently, devoid of interference, thereby enhancing fidelity. The Poincaré sphere is a versatile tool to represent the multi-dimensional parameter space of vectorial light. However, for MDODS beyond three dimensions, it does not account for the essential degree of freedom of light intensity. High-intensity, short-pulsed lasers enable localized two-photon or multiphoton excitation[17,18], providing the basis for the 3D spatial dimensions of MDODS through the X, Y and Z coordinates within the medium.

Therefore, in this study, light intensity ($S_0$ in Stokes space), combined with the vectorial states of light ($S_1$, $S_2$ and $S_3$ in Stokes space), forms a Poincaré-like semi-sphere parameter space (Figure 1b). This multi-dimensional parameter space is a semi-sphere because handedness rarely results in notable material responses in volume light-matter interactions. Thus, a multidimensional storage technology has been developed that utilizes vector light with different parameters within this Poincaré-like semi-sphere to process the medium.

Initially, using vectorial light within this Poincaré-like semi-sphere space to process fused silica glass (Supplementary Information S2), anisotropic nanostructures[11] and chemical bond breakages[19] were induced after laser irradiation. The resulting anisotropic nanostructures exhibit form birefringence, wherein the RET is directly proportional to the degree of structural anisotropy[14]. Their structural orientation is controlled by the azimuthal angle ($\psi$) of the beam due to boundary effects, governed by Maxwell's equations[20]. Additionally, the bond breakages result in defect-induced fluorescence. However, we discovered that the FL and the RET of birefringence are inherently coupled, simultaneously governed by $I_a$ across a wide parameter range ($0 < \text{ellipticity} < 0.7$) （Supplementary Information S2）. This one-to-two correspondence, where a single light parameter dictates two material responses, impedes the independent manipulation of multiplexed signals and leads to inefficient utilization of limited material responses.

The fundamental mechanism underlying this one-to-two correspondence is the wide bandgap of fused silica glass, which requires the absorption of eight photons at a wavelength of 1,030 nm. In higher-order multiphoton excitation, the absorption efficiency of circular polarization is significantly lower than that of linear polarization[22], therefore, changing $I_a$ alters the absorption efficiency, thereby affecting both FL and RET simultaneously. Specifically, governed by Maxwell's equations, boundary effects in the near-field are intricately linked to the vectorial properties of light[20,21]. With consistent light intensity, an increased proportion of circular component diminishes the near-field anisotropy, thereby reducing the RET signal as $I_a$ decreases (Methods, Simulations). FL originates from photochemical processes following MPE. During MPE, circularly polarized light is less efficient than linearly polarized light because of electron dephasing caused by spin angular momentum (SAM) rotational dynamics[22]. As the number of photons ($n$) required for MPE escalates, this dephasing effect becomes more pronounced, thus exacerbating the discrepancy in modifications between circular and linear polarizations[23]. Therefore, in fused silica glass (bandgap ≈ 9.0 eV), linear polarization is associated with higher free carrier density[24], lower laser threshold, and enhanced FL compared to its circular counterpart (Supplementary Information S2). As the portion of $E_b$ increases and $E_a$ decreases, the SAM component rises (Extended Data Fig. 2c), intensifying the dephasing effect and consequently reducing the number of fluorophores generated after MPE. As a result, both RET and FL, which arise from the formation of anisotropic structures, are coupled and modulated by $I_a$ simultaneously (Figure 2a).

**Triple one-to-one correspondence**

Given the intrinsic link between structural anisotropy and the vectorial nature of light, it is crucial to dissociate fluorescence from vectorial states, and align it exclusively with light intensity to establish a multiple one-to-one correspondence. This requires eliminating the influence of the dephasing effect during MPE (Figure 2b). To achieve this, we reduced the number of photons required for processing by incorporating a narrow-bandgap material, thereby suppressing the polarization-dependent absorption and decoupling fluorescence from the vectorial states of light. Specifically, we doped the medium (polyacrylonitrile, PAN) with a narrow-bandgap light absorber ($MAPbBr_3$). The light absorber has a bandgap of 2.25 eV, corresponding to two-photon absorption ($n = 2$) at a wavelength of 1,030 nm (Methods).

Based on the Keldysh theory (Supplementary Information S3)[25], carrier dynamics calculations were conducted for the two material systems with different photon absorption orders. The results show that the dephasing effects in the MPE process caused by SAM are effectively suppressed in two-photon absorption systems (Figure 2f). In the eight-photon absorption system in fused silica, the absorption cross section for linearly polarized light is approximately 3.75 times that for circularly polarized light, leading to a substantially higher carrier density under linear polarization (Supplementary Information Table 1). However, in the two-photon absorption system consisting of $MAPbBr_3$-doped PAN, the differences caused by the absorption cross section are significantly reduced, and the total carrier density under linear and circular polarization is nearly identical (Supplementary Information Table 2).

As a result, the number of free carriers, and consequently the number of fluorophores, is expected to remain nearly constant regardless of the applied elliptical polarization (Figure 2c-f). We achieved a minimal laser-threshold disparity of 4.77% and a fluorescence disparity of 8.64% between linear and circular polarizations in the doped material (Methods, Extended Data Fig. 4a and Fig. 4h), corresponding to only 32% and 4.9%, respectively, of the disparities observed in fused silica (Supplementary Information Fig.S2a and Fig.S2c). Fluorescence from laser-induced carbon dots (CDs) (Extended Data Fig.3a) is uniformly distributed throughout the laser-modified voxel. Ultimately, the FL of the induced modifications depends solely on *I*, irrespective of the vectorial states (Figure 2g and 2g, inset).

Anisotropy is also found within such fluorescence nanostructures (Extended Data Fig. 3f-h). The formation of such anisotropic nanostructures, commonly referred to as 'nanogratings', has been found in many inorganic materials[11,26]. They were fabricated for the first time in bulk organic materials. This indicates that the formation of nanograting does not rely solely on high-temperature or plasma processes specific to inorganic materials, but rather is a universal physical mechanism that can be harnessed. The initial pulse generates an isotropic structure that acts as a seed, promoting near-field anisotropy in subsequent pulse sequences within regions of relatively low temperature[20]. Owing to the emerged anisotropy, the modifications exhibit form birefringence properties. Consistent with expectations derived from the intrinsic link between anisotropic structures and the vectorial properties of light (Figure 2h-k), the birefringence RET of the laser-induced modifications is proportional to $I_a$ (Figure 2l).

By achieving two-photon absorption in organic materials, we have successfully established a double one-to-one correspondence between changes in light parameters and material responses. Building on this, experimental results show that adjusting the azimuth of the incident beam on the Poincaré-like semi-sphere leads to anisotropic field distributions with different orientations (Fig. 2m–p). The orientation of the formed anisotropic nanostructures is perpendicular to the major axis of the polarization ellipse (Extended Data Fig. 3g, h and Extended Data Fig. 10b). Consequently, we introduce a third independent dimension, SAO, as another form birefringence parameter that can be manipulated solely by adjusting $\psi$ (Figure 2q), ultimately establishing a triple one-to-one correspondence (Figure 2b).

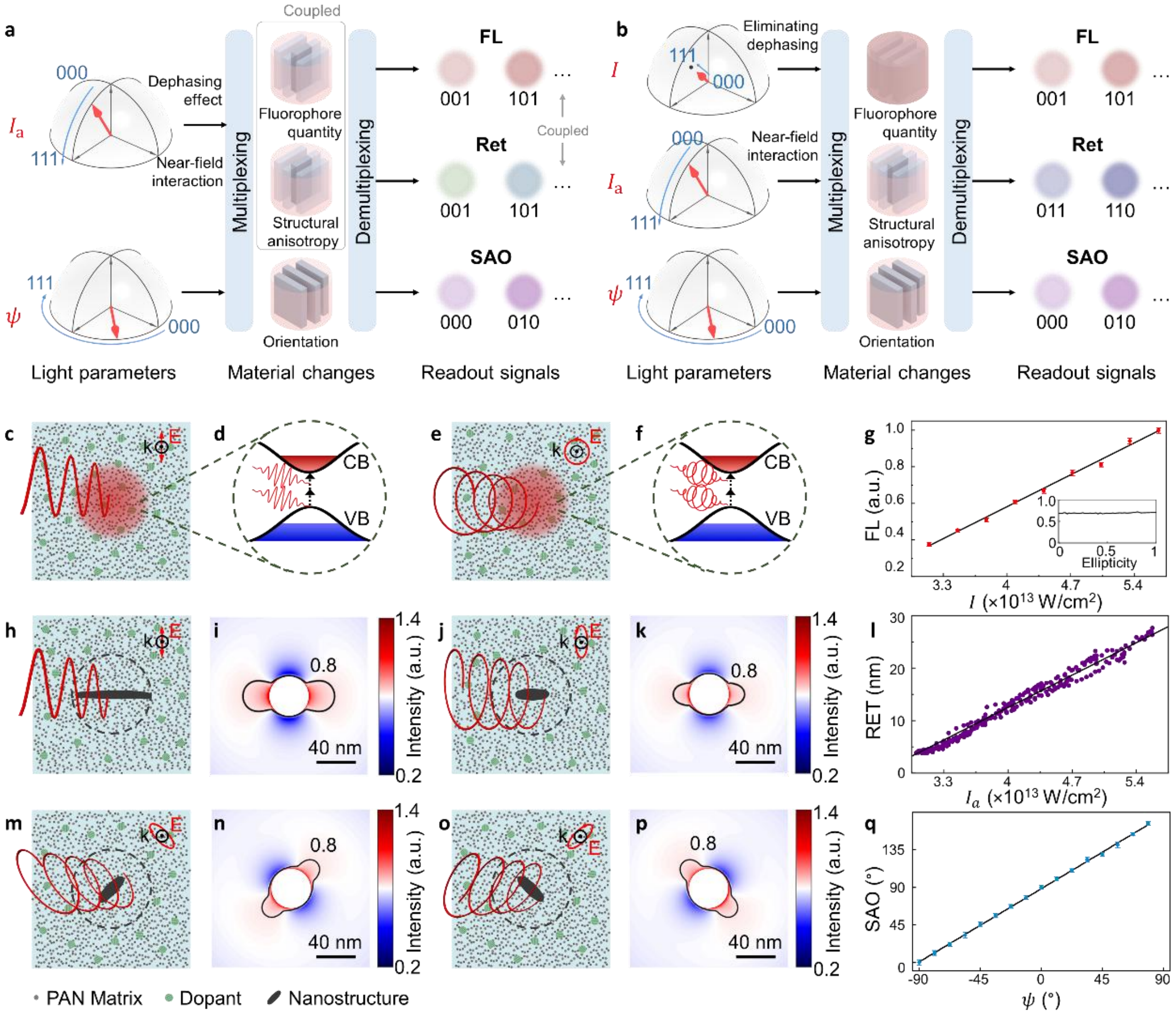

Figure 2 Correspondence between light parameters and material responses. **a**, Illustrates the relationship between light parameters and material responses in fused silica glass. **b**, Illustrates a triple one-to-one correspondence in doped PAN. **c**, **e**, Depict the laser excitation processes using linearly and circularly polarized light, respectively. **d**, **f**, Corresponding schematics of two-photon excitation processes where electrons are excited from the valence band (VB) to the conduction band (CB). **h, j, m, o**, Demonstrate the formation of anisotropic nanostructures induced by varying $I_a$ in (**h, j**) and varying $\psi$ in (**m, o**), respectively. Black dashed circles indicate laser-irradiated areas. Red curves represent electric field components. **i, k, n, p**, Simulations of electric field intensity distribution when light interacts with seed isotropic structures under varying $I_a$ in (**i, k**) and $\psi$ in (**n, p**). Central seed structures are whited-out for better visualization. Solid black lines indicate contour lines of 0.8 intensity, showing near-field anisotropy. **g, l, q**, Graphs showing changes in FL, RET and SAO as functions of $I$, $I_a$, and $\psi$, respectively. The inset in (**g**) presents an example of FL signal at constant $I$ with varying ellipticity.

**High-fidelity 6D multiplexing**

We can now precisely manipulate the FL, RET, and SAO of nanostructures by modulating $I$, $I_a$, and $\psi$ within the Poincaré-like semi-sphere (Figure 1b). Contrary to the one-dimensional subspace created by a linearly polarized beam in the FL-RET space, our approach expands the signal domain to two dimensions (Extended Data Fig. 4i). And introducing SAO as the third orthogonal dimension allows data voxels to possess three degrees of freedom within the FL-RET-SAO space (Extended Data Fig. 4j). Ultimately, by harnessing the 3D space of volumetric materials, we have established an approach of six orthogonal dimensions. Additionally, the 6D multiplexed data voxels are rewritable (Methods).

To validate the viability of 6D multiplexing, we printed the portraits of nine renowned scientists, distributing them across three layers, and encrypting each portrait in the FL, RET, and SAO dimensions (Figure 1d and Figure 3). Each voxel was assigned eight states for FL, four states for RET, and sixteen states for SAO. To create more intuitive images, we opted for evenly spaced values for FL and RET in the FL-RET space. However, for data recording purposes, employing non-uniform signal distributions that fully exploit the FL-RET domain can achieve higher fidelity (Methods).

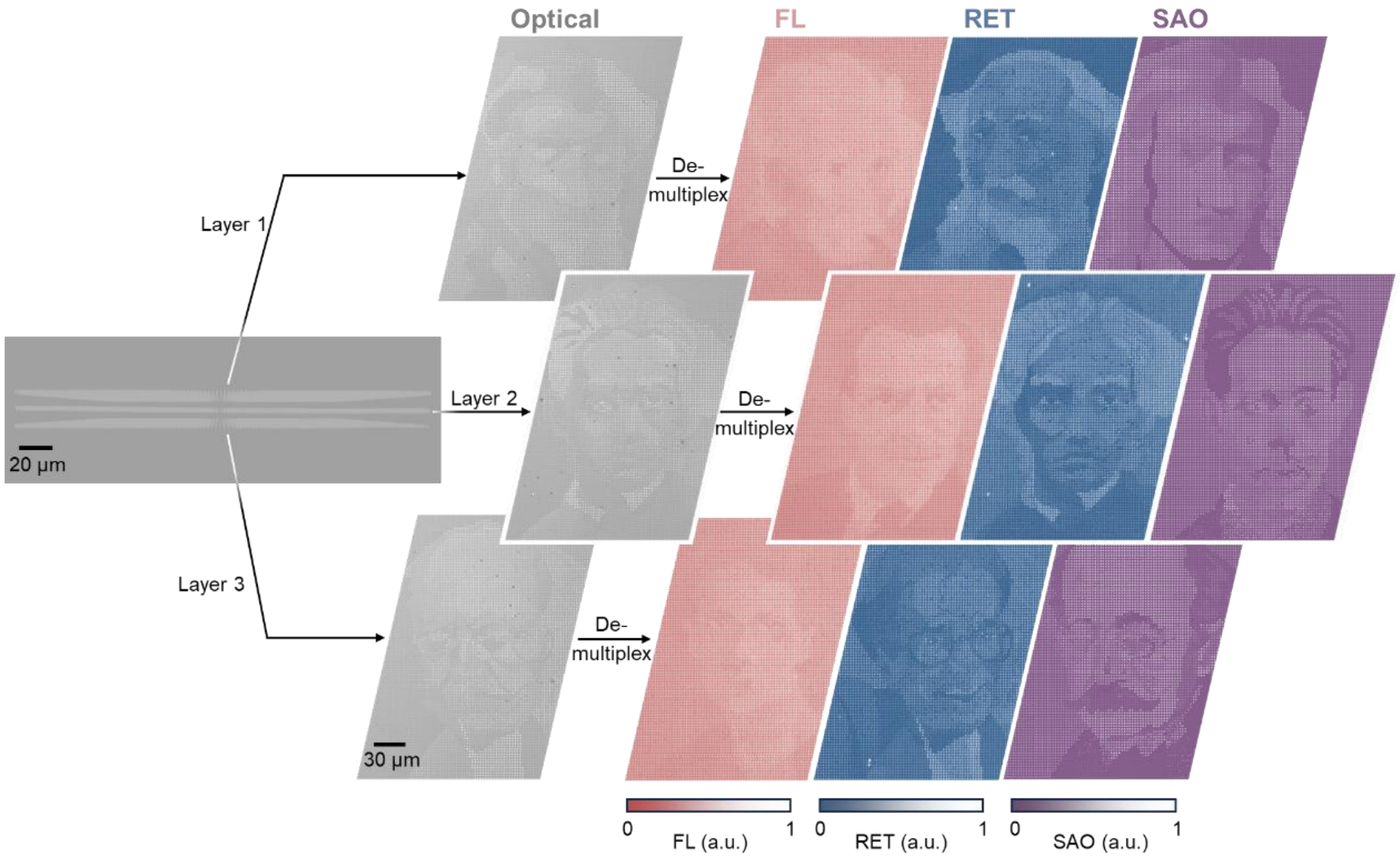

Figure 3 Six-dimensional recording and decoding of multiplexed scientist portraits. A 3D z-stack fluorescence confocal image shows three layers of 100 × 100 voxel arrays. Each voxel array was viewed in optical channel and three multiplexed channels, respectively. The layer separation is 10 μm.

To demonstrate the feasibility of data recording using 6D multiplexing, digital files were written (Figure 4a, and 4a(i)). Raw images of FL, RET, and SAO profiles were obtained, classified, and decrypted (Figure 4b and c). We achieved a raw bit error rate (RBER) of 0.55% (Methods). To visualize the distributions of the multi-dimensional voxels, which encompass 128 states or 7 bits, we employed t-SNE (t-distributed stochastic neighbor embedding) to project them onto a 2D space (Figure 4d)[27]. The resulting t-SNE plot depicts a clear and decentralized clustering distribution, indicating minimal interference among states.

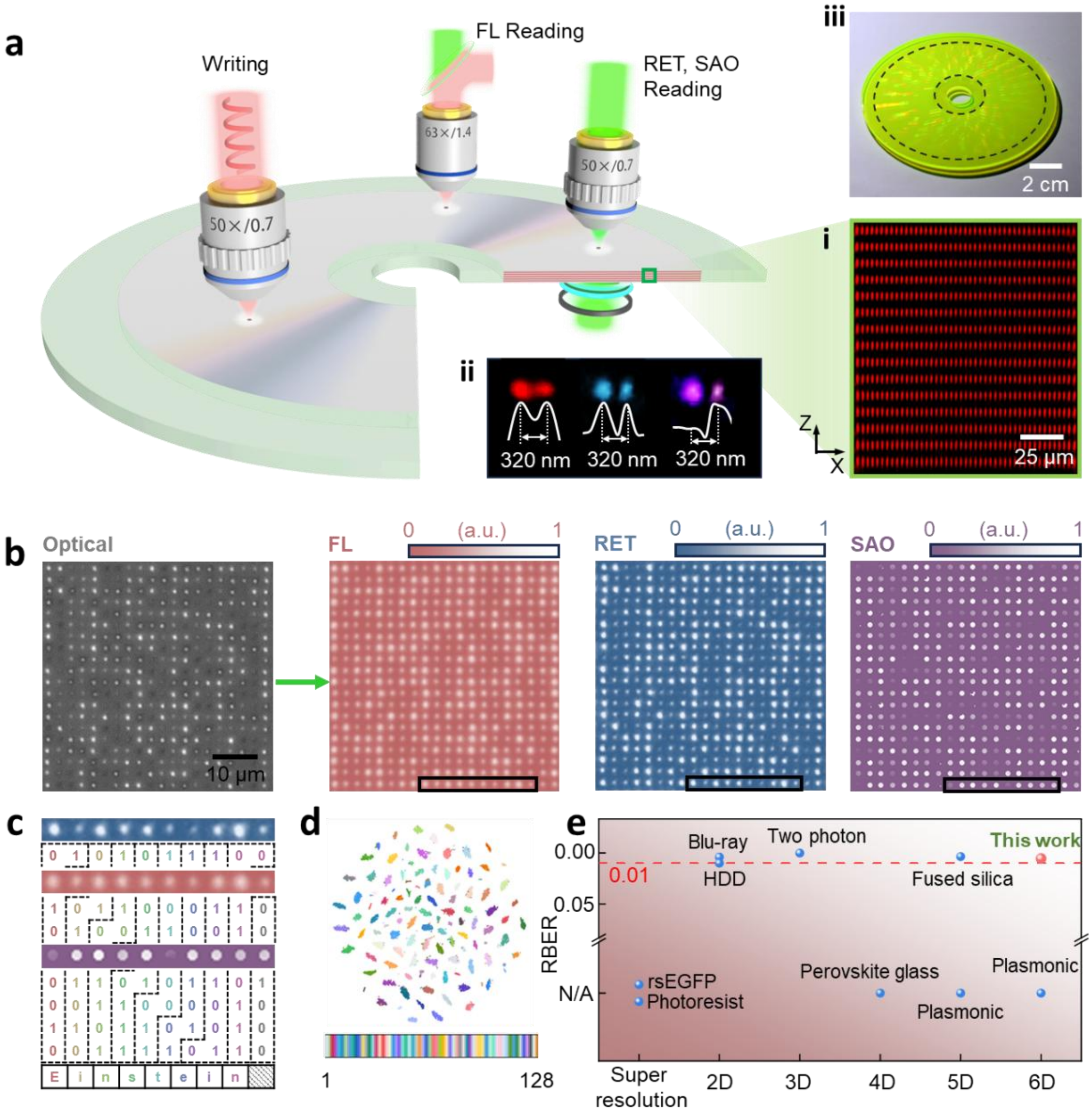

Figure 4 Six-dimensional data recording and decoding. **a**, Schematic of a 6D high-fidelity optical disc with layers written and readout within an isotropic medium. Inset **i** presents an enlarged z-stack fluorescence image of 15 data layers. The layer separation is 10 μm. Inset **ii** shows the FL, RET and SAO profiles of two data voxels separated by 320 nm. Inset **iii** displays a top-view of a 12 cm 6D high-fidelity optical disc after data writing, with the data-written region highlighted by dashed lines. **b**, Illustration of the decoding process, where a multiplexed optical image is demultiplexed into three multi-level profiles: FL, RET, and SAO. **c**, Retrieval process of the eight-byte word 'Einstein' from data voxels highlighted in (**b**), across three dimensions. **d**, t-SNE visualization representing 128 states in 2D space. **e**, Comparison of this work with traditional methods and state-of-the-art techniques regarding RBER[2,5,8,9,13,28–32].

Despite ongoing advancements in next-generation data storage techniques[5,8,9,28,29], the focus on fidelity remains insufficient (Figure 4e). Due to insufficient signal-to-noise ratios, most studies are unable to conduct complete data write-read experiments to obtain reliable raw bit error rates (RBERs). However, it is precisely the RBER that determines the upper limit of a system's coding efficiency at the Shannon limit[7]. Specifically, when RBER exceeds the correction capability of an error correction code, achieving error-free data recording becomes unattainable, even if the code rate approaches zero[7]. The code rate is defined as the ratio of user information bits to the total number of bits generated by the coder. Practically, systems with RBERs exceeding 0.01 are of limited utility due to their diminished code rates. Our approach uses three pairs of one-to-one correspondence to separate the signals during the writing phase, leveraging three light parameters ($I$, $I_a$, and $\psi$) to manipulate three material responses (fluorophore quantity, structural anisotropy, and orientation), each correlating with a distinct physical quantity (FL, RET, and SAO) (Figure 2b), thereby naturally achieving high fidelity. Our approach not only enables 6D optical data storage but also delivers a high-fidelity, coding-efficient write-read system comparable to mainstream practical storage techniques, such as Hard Disk Drives (HDD)[30] and Blu-ray[31]. Additionally, the method achieves a signal-to-noise ratio (SNR) that is an order of magnitude higher than other state-of-the-art techniques[5,9,28] (Extended Data Fig. 5).

**Discussion**

Our 6D technique achieves a minimum dot separation of 320 nm (Figure 4a(ii)). As a proof of concept, we wrote 18 GB of digital data via the 6D technique. Although validated separately (Methods), the combination of 12 bits per voxel, 600 nm voxel separation, and 1 μm layer separation represents a theoretical upper bound of 0.4 petabits (Pb) physical capacity for a standard-sized optical disc (Fig. 4a(iii)). Owing to its inherent high fidelity, our technique proves to be a compelling candidate for real-world applications, particularly in maximizing user capacity. Following the application of proper coding scheme, the user capacity is projected to reach 0.38 Pb (Methods). Currently, the requirement for more than five laser pulses to form each voxel poses a limitation on data writing speed. However, by leveraging techniques such as pulse sequence[33,34] or spatial-temporal modulation[35], we anticipate significant reductions in light deposition times, thereby increasing data writing speeds to GB/s using gigahertz lasers and modulators.

Furthermore, our 6D multiplexing approach has been validated in a variety of other polymers (Supplementary Information S5). Implementing this approach in more durable materials, such as doped glass, could open new avenues for high-fidelity, high-dimensional optical storage. Looking back on the evolution of anisotropic nanostructure-based optical storage from laboratory research towards industrialization attempts[13,36], we anticipate that continued optimization of parameters such as voxel size, interlayer spacing, and write speed[2,12,34–36], will further advance this technology, paving a viable path towards high-capacity optical storage.

## Conclusions

We successfully suppressed the electron dephasing effect during MPE and achieved triple one-to-one correspondence between light parameters and material responses by simultaneously harnessing the principles of photoexcitation and near-field light-matter interactions. Using this approach, we demonstrated high-fidelity six-dimensional optical data storage and validated the feasibility of the write-read process. Furthermore, the photoluminescence tunability across both wavelength[5] and circular dichroism[37] could enable further extension of MDODS into the seventh and eighth dimensions. The creation of multidimensional nanostructures within organic materials could propel advancements in high-capacity data storage, optical elements, microfluidics[38], micromechanics[39], integrated photonics[40], and quantum technologies[41]. More broadly, our strategy of decoupling light parameters from material responses opens new opportunities in fields that require precise control of light–matter interactions.

## Main References:

## Methods

### Visual representation

In the experimental results shown in Figure 2g, nine distinct states of light intensity were used. Each dot on the graph represents the average FL signal from laser-induced voxels at a fixed light intensity but varying ellipticity. For each intensity level, there were 45 different values of ellipticity. In Figure 2l, the RET of the resulting nanostructures is depicted as a function of $I_a$. Voxels displaying a RET lower than 3 nm are excluded because their values are close to the background noise level. Each dot in Figure 2q represents the SAO signal obtained by laser-induced voxels of varying orientation angle. $I$ and $I_a$ were also varied. There were overall 16 states of orientation angle. Error bars in Figure 2g and Figure 2q represent standard deviations. When not visible, the error bars are smaller than the symbol.

The images of fluorescence and retardance in Figure 3 and Figure 4 were directly obtained from raw measurements, but changed the colormap for better visual representation. The SAO dot arrays were extracted from the SAO profiles based on the contours of the dots for intuitive illustration. This is because SAO is non-zero across varying intensities. An example is given in Fig. S1.

### Material fabrication and characterization process

1. **Sample fabrication.** Methylammonium bromide (MABr, 99.9%) and lead bromide ($PbBr_2$, 99.9%) were purchased from Advanced Election Technology Co., Ltd. And N, N-Dimethylformamide (DMF, ≥ 99.5%) was purchased from Sinopharm Group

Co., Ltd. Polyacrylonitrile (PAN, average Mw 150,000) was obtained from Beijing InnoChem Science & Technology Co., Ltd. Fused silica was Corning 7980. All the above reagents were used as received.

A solution of MABr and $PbBr_2$ in a 1:1 molar ratio was dissolved in DMF at 25°C under active mixing for 8 hours to generate a 0.03 M methylammonium lead bromide ($MAPbBr_3$) solution. PAN was then added to the solution and stirred until all visible opaque matter disappeared to obtain a precursor solution with a PAN concentration of 150 mg/mL. The precursor solution was dripped onto a cleaned container to form several millimeters of precursor medium, and residual air bubbles were removed by standing for 12 hours. The precursor samples were then dried in a vacuum oven at 0.1 Pa and 25°C for 24 hours to remove the DMF solution. As the precursor samples dried, they gradually transformed into perovskite quantum dots doped PAN composite medium. Pure PAN samples were fabricated by preparing perovskite-free precursor solutions following a similar strategy.

2. **XRD measurements.** X-ray diffraction (XRD) measurements were performed using a Philips diffractometer (X pert pro-MRD) with a step size of 0.017° and a step time of 10.16 s. The X-ray sources used were Cu $K_{\alpha1}$ and Cu $K_{\alpha2}$, with wavelengths of 1.54060 Å and 1.54443 Å, respectively. The XRD patterns of the perovskite quantum dots-polyacrylonitrile (PQDs-PAN) medium and $MAPbBr_3$ powder were measured and presented in Extended Data Fig. 1a. It is evident that the peaks of PQDs-PAN are in good agreement with that of $MAPbBr_3$ powder. The broadening of the peaks is attributed to the formation of smaller perovskite quantum dots, which is a manifestation of the Scherrer effect. These findings provide strong evidence for the stable presence of PQDs in the PAN matrix.

3. **Transmission electron microscopy measurements.** Transmission electron microscopy (TEM) images were obtained using a field emission high-resolution transmission electron microscope (FEI F200). The samples were embedded and sliced (Leica UC7) before imaging. To investigate the effect of temperature on the size distribution of PQDs, PQDs-PAN samples were prepared at various temperatures and examined their size distributions using TEM, as depicted in Extended Data Fig. 1b. The results revealed a significant increase in the average size of PQDs from $3.78 \pm 0.64$ nm at 290 K to $15.94 \pm 3.92$ nm at 305 K. Higher preparation temperatures led to excessive growth of the PQDs, resulting in agglomeration and reduced stability of the PQDs-PAN medium. Conversely, excessively low preparation temperatures hindered solvent vaporization, impeding sample preparation efficiency. Therefore, considering the balance between medium stability and preparation efficiency, PQDs-PAN mediums prepared at 300 K were selected for further investigation.

4. **Absorption spectra measurements.** The absorption spectra were measured in the 400 ~ 750 nm range using a UV–vis–NIR spectrophotometer (PerkinElmer Lambda 35) in transmission mode with an integrating sphere attachment. The results showed that the band gap of PQDs-PAN was 2.25 eV (Extended Data Fig. 1c). The transmittance (*T*) and absorbance (*A*) values can be mathematically linked through a direct conversion using the formula: $A = -\log_{10}(T)$. The conversion analysis reveals that the transmittance of PQDs-PAN medium (1 mm thickness) exceeds 88% for photons with wavelengths exceeding 550 nm.

## Data writing and readout

1. **Optical configuration for writing.** The experiments were conducted using a mode-locked Yb:KGW femtosecond laser system (Pharos, Light Conversion Ltd.) with a central wavelength of 1,030 nm, a pulse width of 220 fs, and a repetition rate of 100 kHz. A 0.70 NA objective lens with a correction collar was utilized to focus the laser beam for all the experiments. The waveplates were mounted on motorized rotational stages and controlled by a computer program. The samples were placed on an XYZ air-bearing translation stage (Aerotech ABL1000) for parameter sweep experiments to determine the optimal material compositions and laser parameters (Extended Data Fig. 2a). The disc samples were mounted on an air-bearing rotational stage on top of an XYZ translation stage (LAB motion), with the optical path owned similar function (Extended Data Fig. 2a (i, ii, and iii)). A laser pulse sequence technique was employed to increase the writing speed for disc samples[34].

$$\begin{bmatrix}\cos\theta & -\sin\theta\\ \sin\theta & \cos\theta\end{bmatrix}\begin{bmatrix}1 & 0\\ 0 & i\end{bmatrix}\begin{bmatrix}\cos\theta & \sin\theta\\ -\sin\theta & \cos\theta\end{bmatrix}\begin{bmatrix}\cos\left(\frac{\varphi}{2}+\frac{\theta}{2}\right) & -\sin\left(\frac{\varphi}{2}+\frac{\theta}{2}\right)\\ \sin\left(\frac{\varphi}{2}+\frac{\theta}{2}\right) & \cos\left(\frac{\varphi}{2}+\frac{\theta}{2}\right)\end{bmatrix}\begin{bmatrix}-1 & 0\\ 0 & 1\end{bmatrix}\begin{bmatrix}\cos\left(\frac{\varphi}{2}+\frac{\theta}{2}\right) & \sin\left(\frac{\varphi}{2}+\frac{\theta}{2}\right)\\ -\sin\left(\frac{\varphi}{2}+\frac{\theta}{2}\right) & \cos\left(\frac{\varphi}{2}+\frac{\theta}{2}\right)\end{bmatrix}\begin{bmatrix}E_x\\ 0\end{bmatrix}$$

$$= E_x \cdot \begin{bmatrix}\cos\theta & -\sin\theta\\ \sin\theta & \cos\theta\end{bmatrix}\begin{bmatrix}\frac{1}{\sqrt{1+\varepsilon^2}}\\ \frac{i\varepsilon}{\sqrt{1+\varepsilon^2}}\end{bmatrix}$$

The optical path, as depicted in Extended Data Fig. 2a, can be calculated using the Jones matrix. Here, $E_x$ is the electric field of the incident x-polarized beam, $\theta$ is the rotating angle of the quarter-wave plate, $\varepsilon$ is the ellipticity, and $\frac{\varphi}{2}+\frac{\theta}{2}$ is the rotating angle of the second half-wave plate.

Upon applying polarization modulation to the incident beam, the resultant beam exhibits a light intensity equals to $E_x^2$, the azimuthal angle $\psi$ equals to $\theta$ and the ellipticity $\varepsilon$ equals to $\tan(\varphi)$ (Extended Data Fig. 2b).

The elliptically polarized beam can be decomposed into circularly and linearly polarized components (Extended Data Fig. 2c). For a beam of constant intensity, an increase in $E_b$ leads to a corresponding increase in the intensity of the SAM component and a decrease in the intensity of the linearly polarized component.

2. **Optical configuration for reading.** The birefringence measurements were performed using a polarization microscope with a PolScope technique (mounted on an Olympus BX51 microscope)[42]. The 0.70 NA objective lens with a correction collar was utilized. For fluorescence measurements, we employed a confocal microscope system (Zeiss LSM 980) mounted on an inverted microscope. A Plan-Apochromat 63× oil immersion objective lens (1.4 numerical aperture) was utilized for imaging and measurements. During the fluorescence measurements, we set the excitation wavelength to 561 nm, and the detection wavelength ranged from 578 to 723 nm. The fluorescent signal was detected using a photomultiplier tube.

3. **Transmission electron microscopy results of CDs.** For the laser-modified doped PAN, high-resolution transmission electron microscopy (HRTEM) further corroborates the presence of CDs (Extended Data Fig. 3a). Following laser irradiation, intense localized thermal energy facilitates the carbonization of PAN, resulting in the formation of fluorescent CDs (Extended Data Fig. 3b). In contrast, the dopant materials are expected to be decomposed, thereby not contributing to the FL signal.

4. **Photoluminescence spectra and Raman spectroscopy measurements.** Photoluminescence spectra from the laser-treated area reveal a pronounced luminescence peak around 645 nm (Extended Data Fig. 3c). Employing Raman spectroscopy to investigate the photoluminescence's source, the spectra present two peaks at 1,363.7 $cm^{-1}$ and 1,565.8 $cm^{-1}$ (Extended Data Fig. 3d). These peaks correspond with the typical D and G peaks of carbon atoms, respectively[43], indicating the fluorescence originates from CDs formation. Notably, these Raman peaks are evident in the laser-treated PAN, but absent in the laser-treated perovskite, confirming that the CDs stem from the PAN.

5. **Scanning electron microscopy measurements.** The presence of in-volume nanostructures in organic materials creates challenges for their characterizations using conventional polishing and SEM imaging techniques[44]. Mechanical grinding and polishing can damage and distort these delicate organic nanostructures, making them difficult to observe. In this study, we explored alternative methods such as focused ion beam (FIB, Zeiss, GeminiSEM 300) milling and ion beam thinning (Gatan PECS II) on different samples. However, these techniques either failed to modify the medium or destroyed the central regions of the nanostructures due to the significant disparities in ion beam modification thresholds between the laser-modified nanostructures and the pristine medium.

For volume structures, although the central structures within the holes were removed by FIB and ion beam thinning techniques, leaving behind empty holes (Extended Data Fig. 3e), we still observed periodic structures on the inner walls of these holes after milling and tilting (Extended Data Fig. 3f). Interestingly, when the beam was focused several micrometers below the surface, we observed the growth of the structures from inside the material to the surface with an increasing number of pulses (Extended Data Fig. 3g, and h). This growth behavior, also observed in other materials such as fused silica[20] and sapphire[45], offered us a unique opportunity to visualize the nanostructures on the surface under SEM without the need for milling or polishing. These nanostructures, growing from the inside of the material but eventually emerging on the surface, can be regarded as approximate representations of the in-volume nanostructures that could not be directly imaged using SEM techniques.

6. **Processing window and distributions of multiplexed signals.** We conducted a comprehensive analysis of laser-induced modifications in PQDs-PAN. The doped PAN samples were examined using optical microscope to determine the modification threshold. The laser threshold disparity is around 4.77% in doped PAN (Extended Data Fig. 4a). A fluorescence map (Extended Data Fig. 4b) and a retardance map (Extended Data Fig. 4c) were obtained. The fluorescence signal exhibited an increase with higher light deposition, achieved by either accumulating more pulses or increasing pulse energy. However, we observed the formation of voids and the generation of significant stress as the light deposition was further increased. The evolutionary progress of these structures can be intuitively observed in the insets (i, ii, and iii) depicted in Extended Data Fig. 4c.

The structures formed in PQDs-PAN exhibited form birefringence properties similar to those observed in fused silica and other glass materials[2,46,47]. In the experiments where the polarization orientation was varied, the structures were induced using a pulse number of 50 and a pulse energy of 28 nJ (Extended Data Fig. 4d). In the experiments where the pulse energy was varied, the structures were induced using a pulse number of 50 and a polarization orientation of 0° (Extended Data Fig. 4e). Following the measurement of RET and FL, we observed a direct correlation between both signals and pulse energy (Extended Data Fig. 4f). This indicates a bijective relationship between these two physical quantities, suggesting that adjusting pulse energy alone cannot independently control RET and FL.

Voxels were induced by varying pulse energy and ellipticity. It is evident that retardance decreases with increasing ellipticity (Extended Data Fig. 4g). While FL remains nearly constant under the same energy but with varying ellipticity (Extended Data Fig. 4h). The fluorescence fluctuation between linear and circular polarizations is 8.64% (Extended Data Fig. 4h). The achieved 2D distribution of FL and RET, along with the 3D distribution of FL, RET, and SAO values, are visualized in Extended Data Fig. 4i and j.

7. **Signal-to-noise ratio.** Most studies focusing on next-generation optical data storage failed to report the RBER. Some obtained the RBERs based on images which encoded predominantly with replicate or consecutive states, rather than real digital files, rendering such RBER measurements unreliable. Similarly, SNRs are seldom disclosed.

We define the following parameters for SNR analysis:

For two-state encoding:

- $A_i$: Sample value for state 1.
- $A_j$: Sample value for state 2.

- $M_1$: Mean of the signal of state 1, $M_1 = \frac{\sum A_i}{n}$.
- $M_2$: Mean of the signal of state 2, $M_2 = \frac{\sum A_j}{n}$.
- $S$: Signal strength, $S = (M_1 - M_2)^2$.
- $N_1$: Noise intensity: $N_1 = \frac{\sum (A_1 - M_1)^2}{n}$.
- $N_2$: Noise intensity: $N_2 = \frac{\sum (A_j - M_2)^2}{n}$.
- $N$: Noise intensity: $N = (N_1 + N_2)/2$.
- *SNR=S/N.*

In multi-state encoding, the overall SNR is calculated as the average of the SNRs for each pair of adjacent states. From the results presented in Extended Data Fig. 5, it is evident that the SNR achieved in this work is an order of magnitude higher than that reported in other studies[5,9,29]. This high SNR and fidelity have enabled us to write real digital data on a relatively large scale, thus facilitating the accurate calculation of RBER.

8. **Portraits painting.** The signal domain generated by 6D multiplexing exhibited a trapezoidal shape in two dimensions (Extended Data Fig. 4i). To facilitate the portrait printing, we selected FL and RET signals with evenly spaced values, starting from an identical initial value (Extended Data Fig. 6a). This enabled us to encode and intuitively print the portraits of scientists using the FL and RET signals. We expanded our encoding scheme by introducing the SAO as the third dimension, enabling the encrypting of each voxel matrix into three separate channels. This enabled us to print nine scientist portraits from three staggered layers (Extended Data Fig. 6c).

**9. Six-dimensional writing and capacity.** The uniform distribution of FL and RET did not fully utilize the entire area of the two-dimensional signal domain (Extended Data Fig. 6a). The values were not sufficiently separated in the FL-RET space, which could potentially lead to decoding errors. To overcome this limitation, we employed a non-uniform distribution of FL and RET that closely matched the shape of the signal domain (Extended Data Fig. 6b). A similar distribution as in Extended Data Fig. 6b was employed in our data recording experiments to achieve high readout fidelity.

We have successfully demonstrated data multiplexing by storing eight, four, and sixteen states of grey-scale scientist portraits in FL, RET, and SAO, respectively. This results in a total of 9 bits of information per voxel. Additionally, we demonstrated that 12 bits of information per voxel can be stored with a 600 nm separation (Extended Data Fig. 7). These 12 bits correspond to 4,096 states for multi-level recording through 16, 8, and 32 states of FL, RET, and SAO, respectively (Extended Data Fig. 7a, b, and c). A series of experiments confirmed that voxel matrices with a 600 nm separation maintain high-quality results (Extended Data Fig. 7h). Furthermore, through analyzing birefringence profiles and atomic force microscopy (AFM) measurements after focused ion beam (FIB) drilling, we demonstrated that multi-layer data voxel arrays are staggered longitudinally with an axis separation of 1 μm (Extended Data Fig. 7d-g).

Consequently, employing a 600 nm voxel separation, a layer separation of 1 μm and 12 bits per voxel, a standard 4.7-inch diameter, 1.2 mm thick 6D optical disc can achieve a storage capacity of 0.4 Pb. Looking ahead, we anticipate that advancements such as high numerical aperture objective lenses, shorter wavelengths, and shingled writing techniques could further enhance the storage capacity[20,47].

We also developed a disc library tailored for our 6D storage system (Extended Data Fig. 8). The library is 0.51 m × 0.54 m × 1.03 m and can store up to 2,200 discs. Assuming each disc has a capacity of 0.4 Pb, the total data storage per library is 880 Pb, achieving a data density of approximately 3.03 $Eb/m^3$. Video demonstrating the handling and movement of the 6D high-fidelity discs within the library can be found in Supplementary Video 2. The outer packaging of the library was removed to facilitate easier observation of disc movement.

10. **Data voxel rewriting.** The data rewriting can be achieved by changing the azimuthal angle in the polarization ellipse of the rewriting laser beam. As a proof of concept, the Extended Data Fig. 9a and b show the change of RET and SAO with varying pulse numbers of the rewriting beam. To simplify the presentation of the rewriting experiment, we demonstrated the rewriting process using orthogonal polarizations. As the number of rewriting pulses increases, the retardance initially reduces and then increases (Extended Data Fig. 9a), and the slow axis re-orients after approximately 15 pulses (Extended Data Fig. 9b). These results indicate the erasure and rewriting of the voxels. Based on these characteristics, both the retardance and slow axis of the birefringent data voxels can be rewritten (Extended Data Fig. 9c). The fluorescence signal increases as the number of pulses increases, as a result of the accumulation of fluorescent CDs through continuous energy deposition. The rewriting of the fluorescence signal was achieved by increasing the fluorescence to pre-defined levels and subsequently reclassifying them. To illustrate this process, two dot arrays with different fluorescence levels were initially written, and they were subsequently rewritten using varying pulse numbers (Extended Data Fig. 9d). Voxels with a lower FL level (State 0) can be rewritten to either a low intensity level (State 0 (rewritten)) or a high intensity level (State 1 (rewritten)) by changing the number of pulses of the rewriting beam. And similarly, voxels with a higher FL level (State 1) can be rewritten to either a low intensity level (State 0 (rewritten)) or a high intensity level (State 1 (rewritten)). The rewriting process can be conducted for three cycles with no serious fluctuations (Extended Data Fig. 9f and g). Although the rewriting cycles are not unlimited, they are sufficient for warm/cold data storage applications.

11. **Data decoding and RBER.** Multiple matrices of 6D data were recorded in PQDs-PAN. Each matrix consisted of 200 × 200 voxels, with each voxel containing 7 bits of data, resulting in a total data capacity of 35 kB per matrix (Extended Data Fig. 6d). To decode the six-dimensionally encoded data, we utilized a simple and lightweight convolutional neural network (CNN). The network architecture is illustrated in Extended Data Fig. 6e.

The dataset was divided into three subsets: training, validation, and testing using a ratio of 4:1:1. The input images provided to the network consisted of separate channels for FL, RET, and SAO. The decoding bit accuracy was calculated as an average of five runs, resulting in an average RBER of 0.55% for a 255.5 kB data.

12. **User capacity retention.** In practical data storage systems, the actual user capacity is determined by several factors, including medium capacity, RBER, and the coding scheme employed. It is widely agreed upon that the RBER should be below $10^{-1}$ to allow for subsequent modulation and error-correction coding in order to meet application-level error-tolerant standards (e.g., $10^{-12}$), albeit at the cost of storage efficiency. Specifically, the RBER needs to be below 7.89% to ensure bit-error correctable information storage[49], as determined by the Shannon limit. Nevertheless, having a high RBER often necessitates the use of longer codewords and additional parity bits, leading to a significant reduction in the code rate. The code rate, which serves as a measure of storage efficiency, is defined as the ratio of the number of user information bits to the total number of generated bits by the coder.

The complexity of the storage system increases because the space in the storage medium is utilized more efficiently. This increased complexity gives rise to mutual influences among the data storage voxels, leading to an increase in the SNR, which is directly related to the RBER. If the RBER cannot be effectively controlled to a reasonable value, the cost of correcting the data will surpass the benefits gained from increased capacity. Therefore, achieving a low RBER is crucial to maintain the overall effectiveness and practicality of a high-capacity data storage system.

Here, we analyzed the relationship between the RBERs and their corresponding code rates using an error correction toolbox[49] and widely used Low-Density Parity-Check (LDPC) codes[50]. The plotted spots represent the RBERs and code rates required to achieve an overall bit error rate of $10^{-5}$ using standard LDPC coding (Extended Data Fig. 6f). Two LDPC codes were plotted for demonstration purposes. We used a long LDPC code with a codeword length of 12,288, and a short LDPC code with a codeword length of 1,536. We selected three code rates for the experiment, 1/2, 2/3, and 4/5, respectively. The RBER for the Blu-ray technique was $4.2\times10^{-3}$, determined based on the Random-Symbol Error Rate after Partial-Response-Maximum-Likelihood[31]. The actual RBER is expected to be higher than this value. For HDD, the RBER is expected to be lower than $1\times10^{-2}$ to ensure that user information is accurately recovered[29].

The results obtained from LDPC coding demonstrate a clear trend: as the RBER increases, the code rate decreases sharply (Extended Data Fig. 6f). For example, with long LDPC coding, the code rate starts at 4/5 when the RBER is $1.33 \times 10^{-2}$, but decreases to 1/2 when the RBER reaches $4 \times 10^{-2}$. Notably, as the RBER approaches $7 \times 10^{-2}$, the code rate experiences a significant drop, indicating that the system (when long LDPC is applied) becomes uncorrectable.

In our work, we have achieved a low RBER (0.55%), allowing for a code rate close to one (0.961 based on the fitting curve). This high code rate ensures maximum retention of user capacity after modulation is applied. By maintaining a low RBER, we can preserve a significant amount of information while minimizing data loss, thereby optimizing the performance of our data storage system. The estimated user capacity is 0.38 Pb.

**Simulations and discussions**

1. **Multiplexing that relies on single material change.** Reliance on a single material change introduces interference during the data writing and readout processes, consequently reducing fidelity. In the writing stage, the employment of only one type of material change necessitates that the material possesses distinctive absorption properties responsive to various light parameters, such as wavelength, polarization, and orbital angular momentum. Nevertheless, in plasmonic-based multi-dimensional data storage systems, despite the unique structural characteristics of the nanostructures, they often exhibit partially overlapping absorption spectra[8,51]. This overlap leads to spectral interference, presenting significant challenges to the practical robustness of the writing process.

Furthermore, employing only a single type of material change necessitates packing nano-materials with different states into a confined voxel space. This approach is manageable when the material change occurs at a molecular scale, such as in holographic optical memory, where numerous molecules can be modulated within a voxel-sized space. However, complications arise when the material change occurs on a scale of tens of nanometers. For instance, in plasmonic data storage schemes, the spatial constraints within each diffraction-limited voxel significantly limit the number of nanostructures, such as nanorods. This scarcity of distinct nanostructures severely undermines fidelity, especially as the number of multiplexed dimensions increases.

During readout, while a range of light parameters are utilized for multiplexing, they influence only a single material property. Consequently, all physical quantities are derived from this singular property and manifested as readout signals. This convergence leads to severe signal interference during readout, substantially hampering the clarity of the output signals.

2. **Simulations.** In previous studies, both the finite-element method and finite-difference time-domain (FDTD) techniques have been employed to simulate near-field anisotropy processes, assuming seed structures as either nanovoids or nanoplasmas[20,21,44]. The results have been consistently similar, which is expected given that the near-field anisotropy process is predominantly governed by boundary conditions. Regardless of the type of modification-be it plasma, self-trapped excitons, defects, fluorophores, or structural changes—the boundary defined by refractive index differences persists. These intermediate or permanent modifications, exhibiting different refractive indices compared to pristine materials, are collectively referred to as ‘seed modifications’ under both single and multi-pulse irradiation conditions.

For simplicity, these seed modifications are assumed as nanovoids induced by the initial laser pulses. When the incident beam is circularly polarized, there is no difference in light intensity at different boundaries (Extended Data Fig. 10a). However, as $I_a$ increases, the disparity in intensities at the boundaries becomes more pronounced, reaching a maximum when the light becomes completely linearly polarized (Extended Data Fig. 10c). This anisotropy in the light field induces plasmas that mirror the shape of the light field distribution, resulting in permanent structural changes. The orientation of the anisotropy is perpendicular to $\psi$ (Extended Data Fig. 10b). The degree of anisotropy in these structures correlates directly with the optical retardance, which escalates with increasing $I_a$. Thus, the value of retardance drops as $I_a$ decreases. Although these simulations are conducted using PAN as the medium, similar outcomes are expected when the substrate is fused silica, due to the universal principles of boundary conditions.

3. **Applicability of the 6D multiplexing**. The signal domain created by our 6D technique exhibits a tobacco pipe-like geometry within silica glass. This observation suggests that doped glasses with high robustness could potentially serve as a high-capacity medium for 6D eternal data storage (Supplementary Information S2). Moreover, we have demonstrated that polyvinyl butyral and polyvinylpyrrolidone are viable candidates for 6D data storage, in addition to PAN (Supplementary Information S4). Furthermore, beyond $MAPbBr_3$, our results indicate that formamidinium lead bromide ($FAPbBr_3$) with a bandgap of 2.30 eV is also suitable as a dopant (Supplementary Information S4).

We anticipate that extending the 6D data high-fidelity storage technique to other materials will follow a methodology akin to what has been demonstrated with doped PAN. This approach aims to generate fluorescence anisotropic nanostructures exhibiting both fluorescent and birefringent properties (Extended Data Fig. 10d), with the decoding approach mirroring our established method (Extended Data Fig. 10e).

**Data availability**

The data that support the findings of this study are available at https://doi.org/10.57760/sciencedb.17084.

**Code availability**

The MATLAB code used in this study is available at https://doi.org/10.57760/sciencedb.17084.

**Methods References:**

**Acknowledgements:**

This work was supported by National Key Research and Development Program of China (2024YFB4505100), Creative Research Group Project of NSFC (61821003), Hubei Provincial Natural Science Foundation of China (2026AFA116), Innovation Fund of the Wuhan National Laboratory for Optoelectronics, Program for HUST Academic Frontier Youth Team and Innovation Project of Optics Valley Laboratory. We are grateful to J. Qiu from Zhejiang University, X. Li from Jinan University for reviewing and discussing the first draft of our manuscript.

**Competing interest declaration:**

The authors declare no competing interests.

**Additional information:**

**Supplementary Information** is available for this paper.

**Correspondence and requests for materials** should be addressed to Jingyu Zhang, Guangda Niu and Hong-Bo Sun.

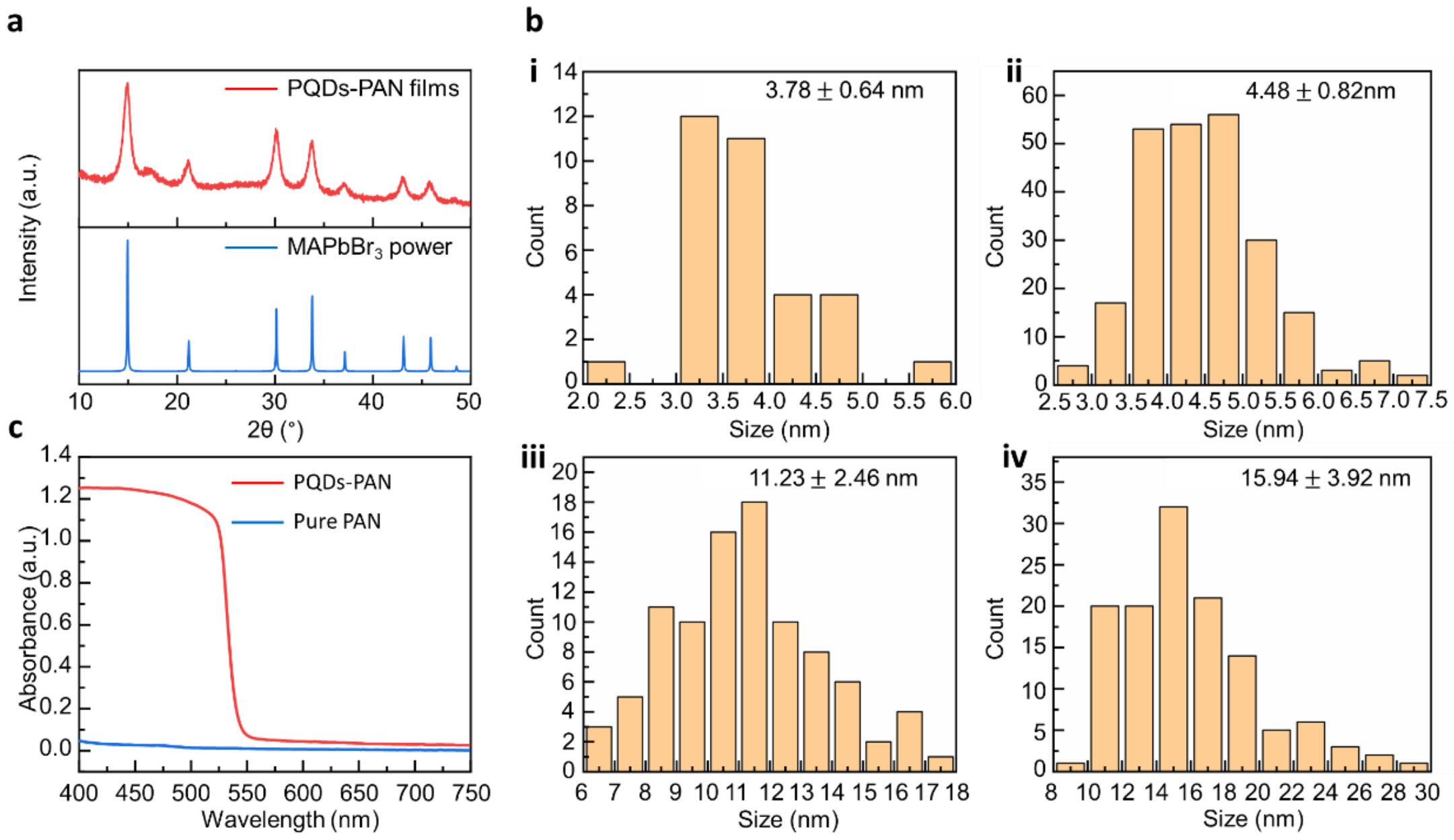

Extended Data Fig. 1 **a,** XRD spectrum of the PQDs-PAN medium and perovskite ($MAPbBr_3$). **b,** Size distributions of perovskite quantum dots synthesized at different temperatures. (**i**) 290 K. (**ii**) 295 K. (**iii**) 300 K. (**iv**) 305 K. **c,** The absorption spectra of the PQDs-PAN sample and pure PAN sample.

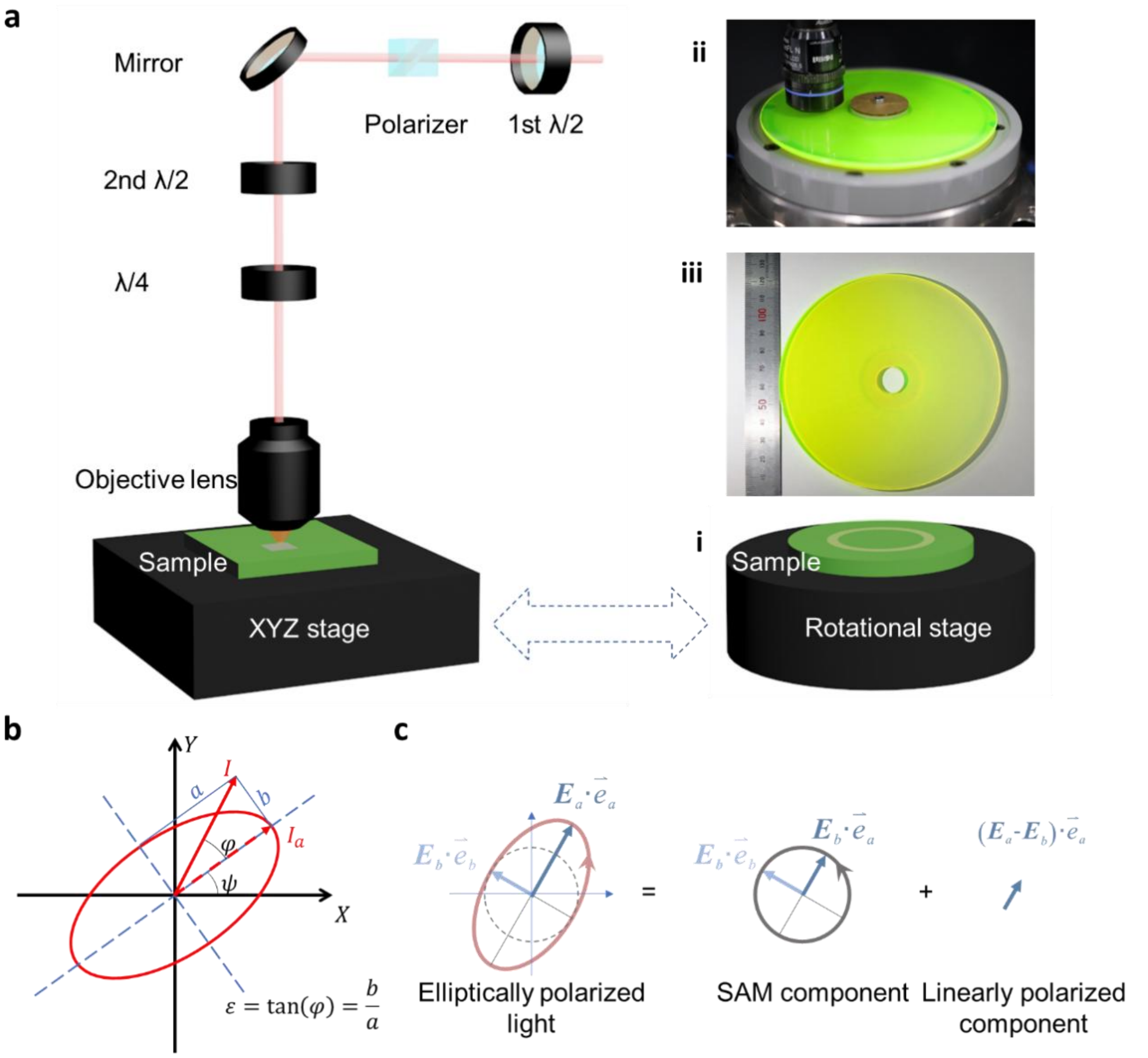

Extended Data Fig. 2 **a,** Schematic of the experimental setup with an XYZ translation stage. (**i**) Schematic of the rotational stage. (**ii**) Recording data on a PQDs-PAN optical disc placed on an air-bearing rotational stage. (**iii**) Top view of the PQDs-PAN optical disc. **b,** Schematic of elliptically polarized light. **c**, Schematic representation of elliptically polarized light segmentation into circularly and linearly polarized components.

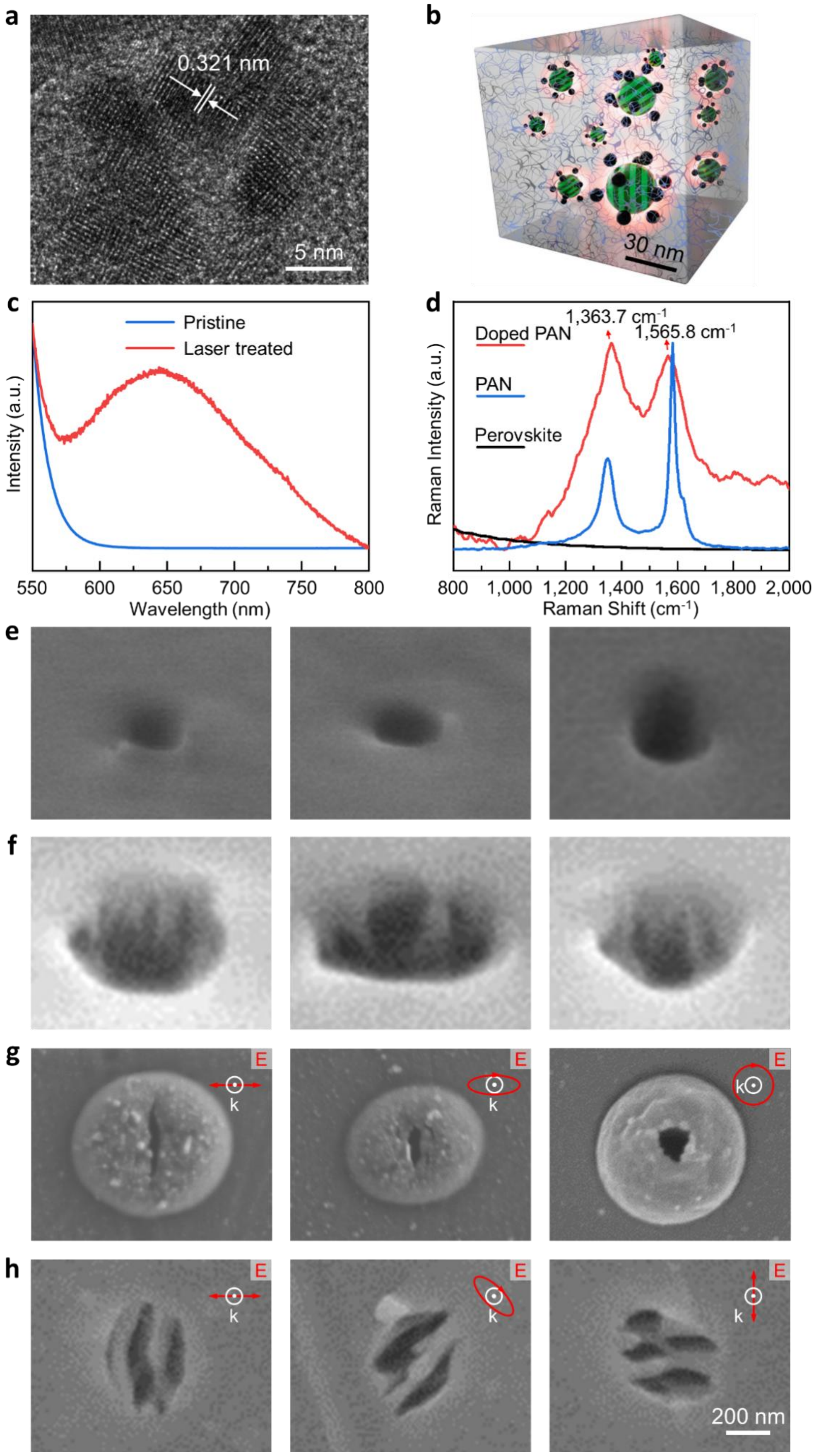

a
0.321 nm
5 nm
b
30 nm
c
Pristine
Laser treated
Intensity (a.u.)
550
600
650
700
750
800
Wavelength (nm)
d
1,363.7 cm-1
1,565.8 cm-1
Doped PAN
PAN
Perovskite
Raman Intensity (a.u.)
800
1,000
1,200
1,400
1,600
1,800
2,000
Raman Shift (cm-1)
e
f
g
E
k
h
E
k
200 nm

Extended Data Fig. 3 **a,** HRTEM image reveals CDs in the laser-treated doped PAN. **b,** Schematic enlarged view from illustrating PQDs (green dots), CDs (black dots), and PAN polymer (random curves). **c,** Photoluminescence spectra comparing the doped PAN medium before and after laser treatment. The spectra was measured using a LabRAM HR800 with the excitation wavelength set to 532 nm. **d,** Raman spectra for laser-treated doped PAN, PAN, and perovskite crystal. A laser confocal Raman spectrometer (Horiba Jobin-Yvon LabRAM HR800) was used. **e**, SEM images of the nanostructures after FIB milling. **f**, SEM images of the nanostructures after FIB milling and tilting. **g, h,** Surface SEM images of the top of the nanostructures without any milling or polishing.

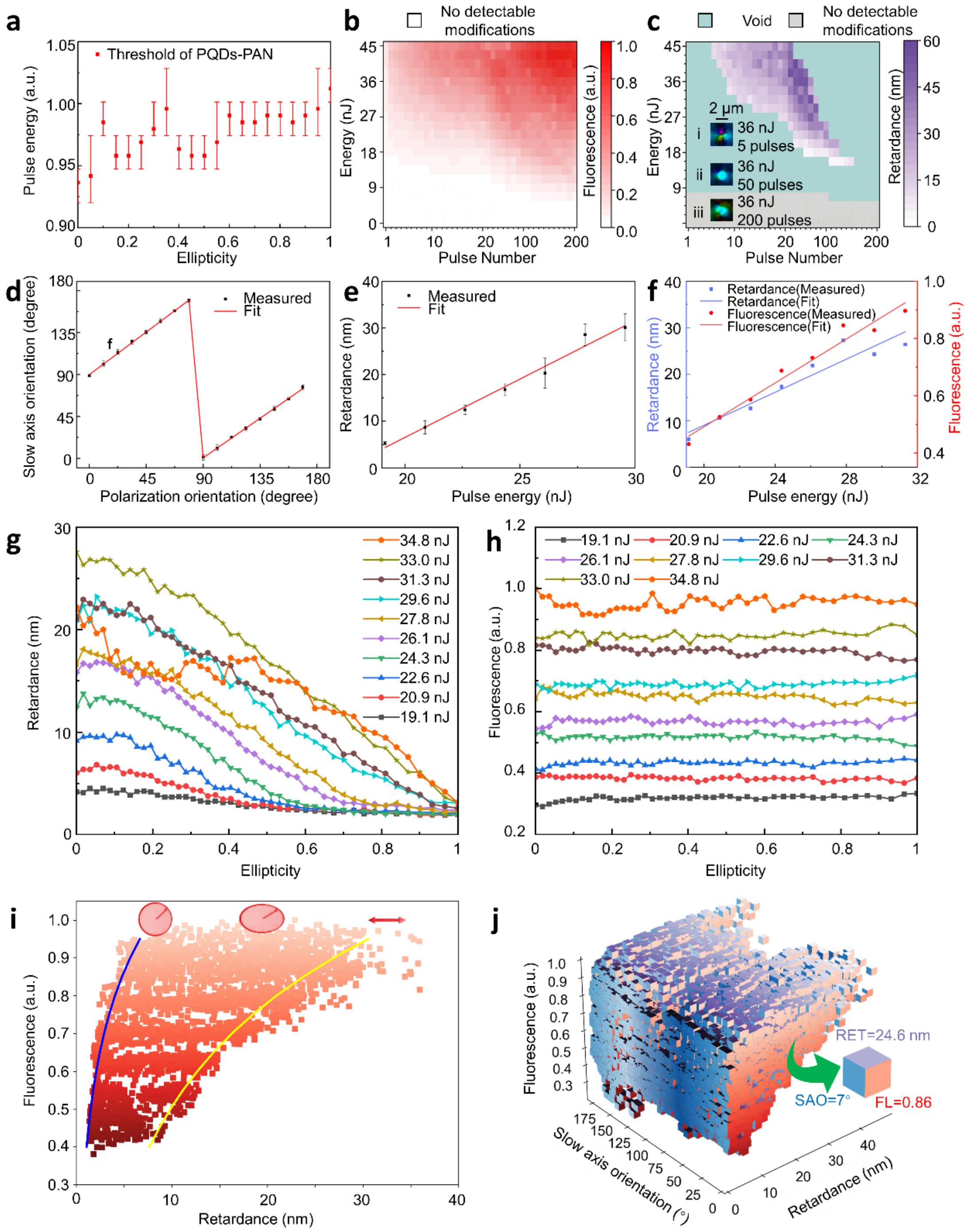

Extended Data Fig. 4 **a**, The single pulse modification threshold of doped PAN as a function of ellipticity. **b**, Fluorescence measurements of laser-induced modifications in PQDs-PAN with varying pulse number and pulse energy, using a linearly polarized light. **c**, Retardance measurements of laser-induced modifications in PQDs-PAN, obtained by varying the pulse number and pulse energy of a linearly polarized light. Structures exhibiting uniform birefringence profiles are categorized as nanogratings (as shown in inset ii). Structures exhibiting randomly distributed birefringent signals are categorized as voids (as shown in inset i and iii). The contrast of the profile in inset i was enhanced

to improve visibility. **d**, Measured slow axis orientation as a function of polarization orientation. **e**, Measured retardance as a function of pulse energy. **f**, Retardance (left) and fluorescence (right) as a function of pulse energy for a linearly polarized light. The pulse number was 50 and the polarization orientation was 0°. **g**, Retardance and **h**, Fluorescence as a function of ellipticity and pulse energy. **i**, 2D distribution of FL and RET, obtained by varying ellipticity and laser intensity, while maintaining a constant polarization orientation at 0°. Blue and yellow curves represent fits from data corresponding to circular and linear polarizations, respectively. **j**, Visualization of FL, RET, and SAO values scattered within a 3D virtual space. Each cube represents a distinct data voxel created within the light parameters in the Poincaré-like semi-sphere. The colors and positions of the cubes correspond to their specific values. A representative cube is magnified in the inset for better illustration.

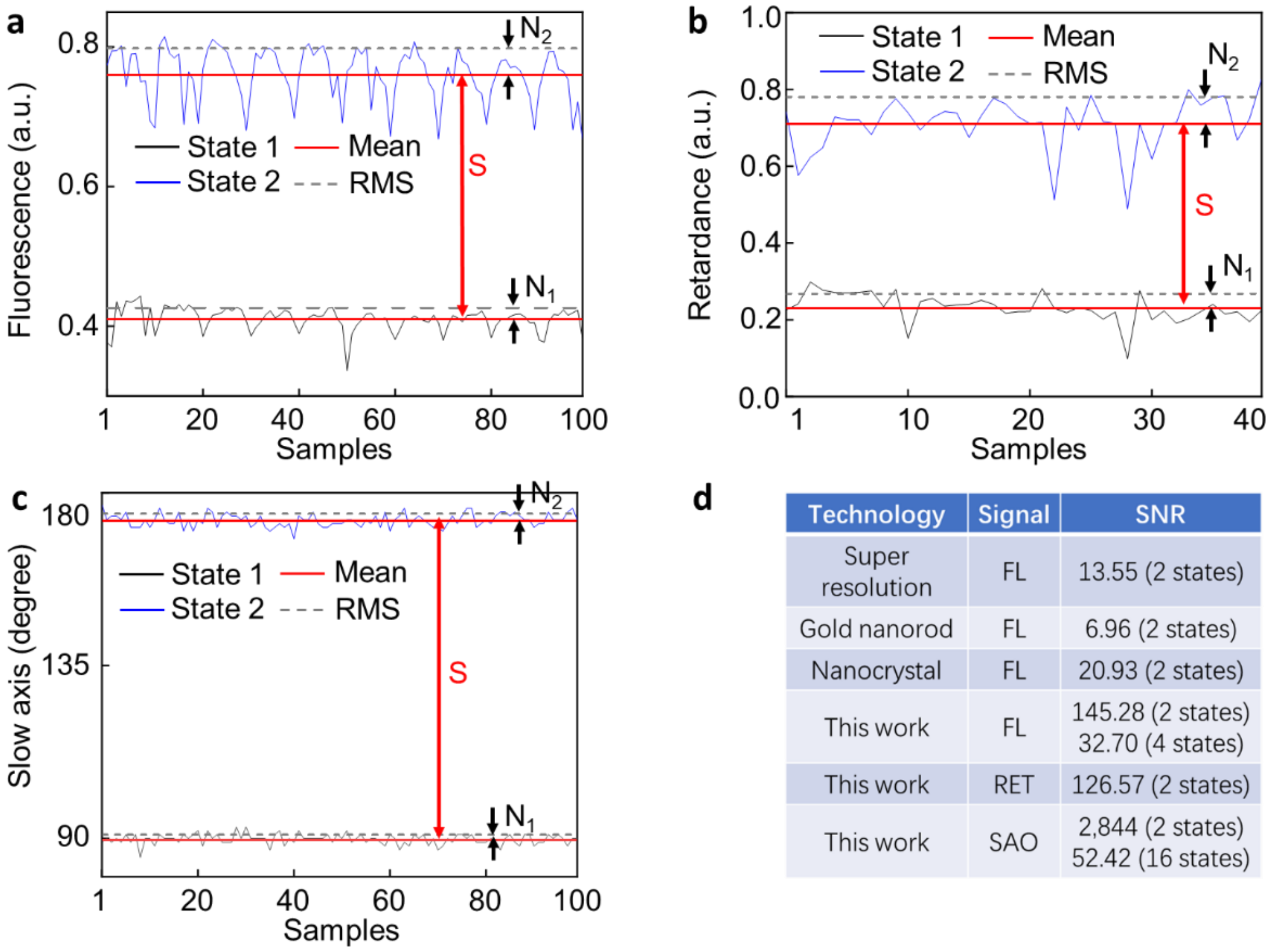

| Technology | Signal | SNR |
|---|---|---|
| Super resolution | FL | 13.55 (2 states) |
| Gold nanorod | FL | 6.96 (2 states) |
| Nanocrystal | FL | 20.93 (2 states) |
| This work | FL | 145.28 (2 states)<br>32.70 (4 states) |
| This work | RET | 126.57 (2 states) |
| This work | SAO | 2,844 (2 states)<br>52.42 (16 states) |

Extended Data Fig. 5 Examples of SNR calculations for FL, RET, and SAO are provided in **a**, **b**, and **c**, respectively. **a**, FL signals at two distinct states. **b**, RET signals at two distinct states. **c**, SAO signals at two distinct states. The numbers of samples are limited for brevity. **d,** A comparison of SNRs among various technologies is provided. The SNRs from other studies were calculated based on data voxel profiles within the images and their corresponding color bars[5,9,29].

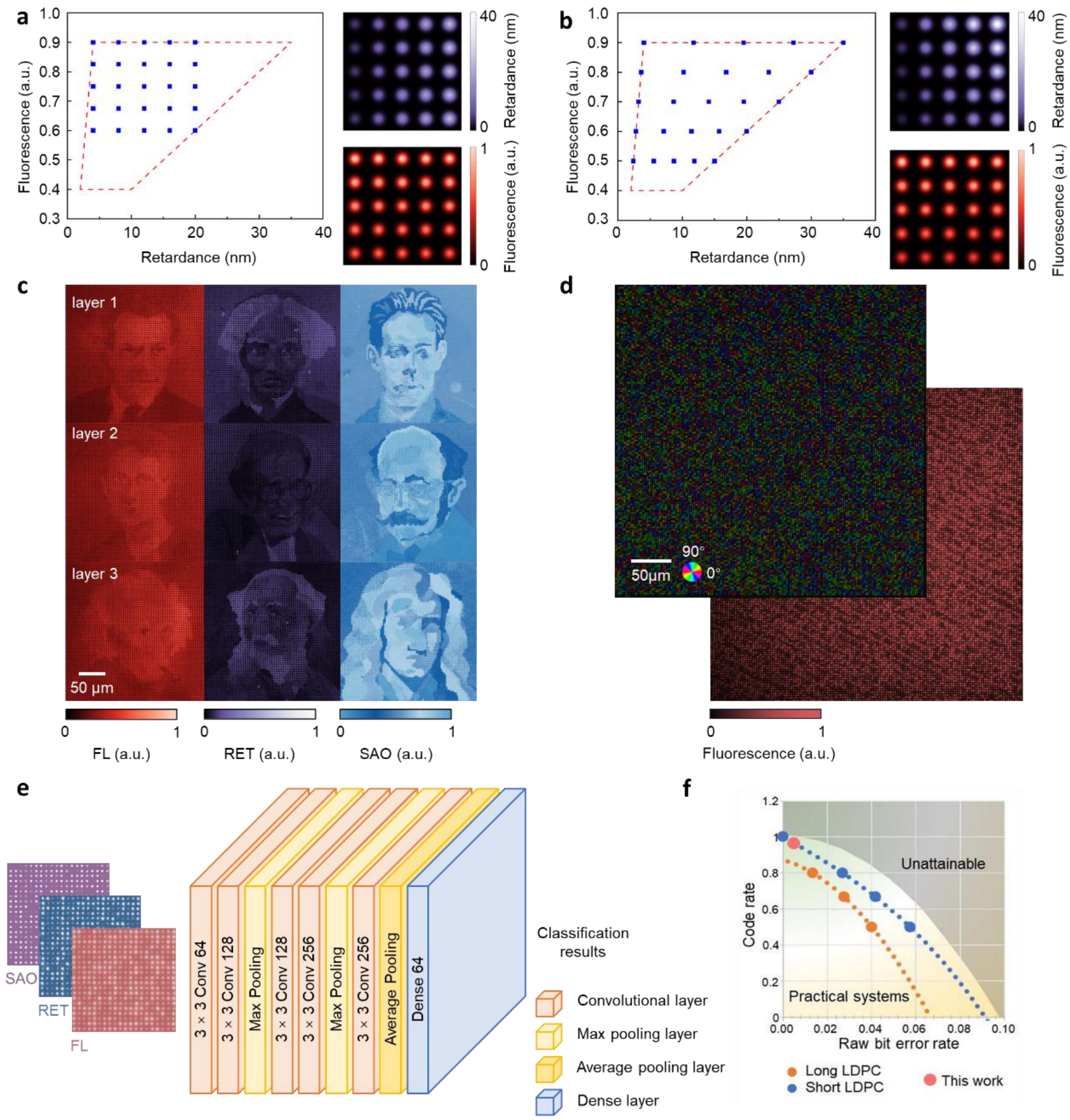

Extended Data Fig. 6 **a, b**, 5 × 5 matrixs illustrating (**a**) evenly spaced and (**b**) non-uniformly distributed fluorescence and retardance values. Each blue square corresponds to a group of voxels with corresponding retardance and fluorescence values. The dashed trapezoidal shape represents the created signal domain. Insets provide schematic views of the corresponding retardance and fluorescence profiles as in (**a**) and (**b**), respectively. **c,** Nine scientist portraits encoded by FL, RET and SAO signals in three staggered layers. Each portrait contains 100 × 100 voxels. **d,** Example of the readout data matrix captured under a birefringence microscope (left) and a fluorescence microscope (right). The birefringence image includes both retardance and slow axis orientation information, represented by transparency and colors, respectively. Pseudo colors in inset indicate the slow axis orientations. **e**, Illustration of the architecture of the designed CNN. **f**, The code rate of our 6D technique, with the assumption that two LDPC codes are used. In line with Shannon's coding theorems, the unattainable area, depicted here for illustrative purposes, signifies RBERs and code rates where reliable data storage is unfeasible. Dashed lines denote the corresponding fits.

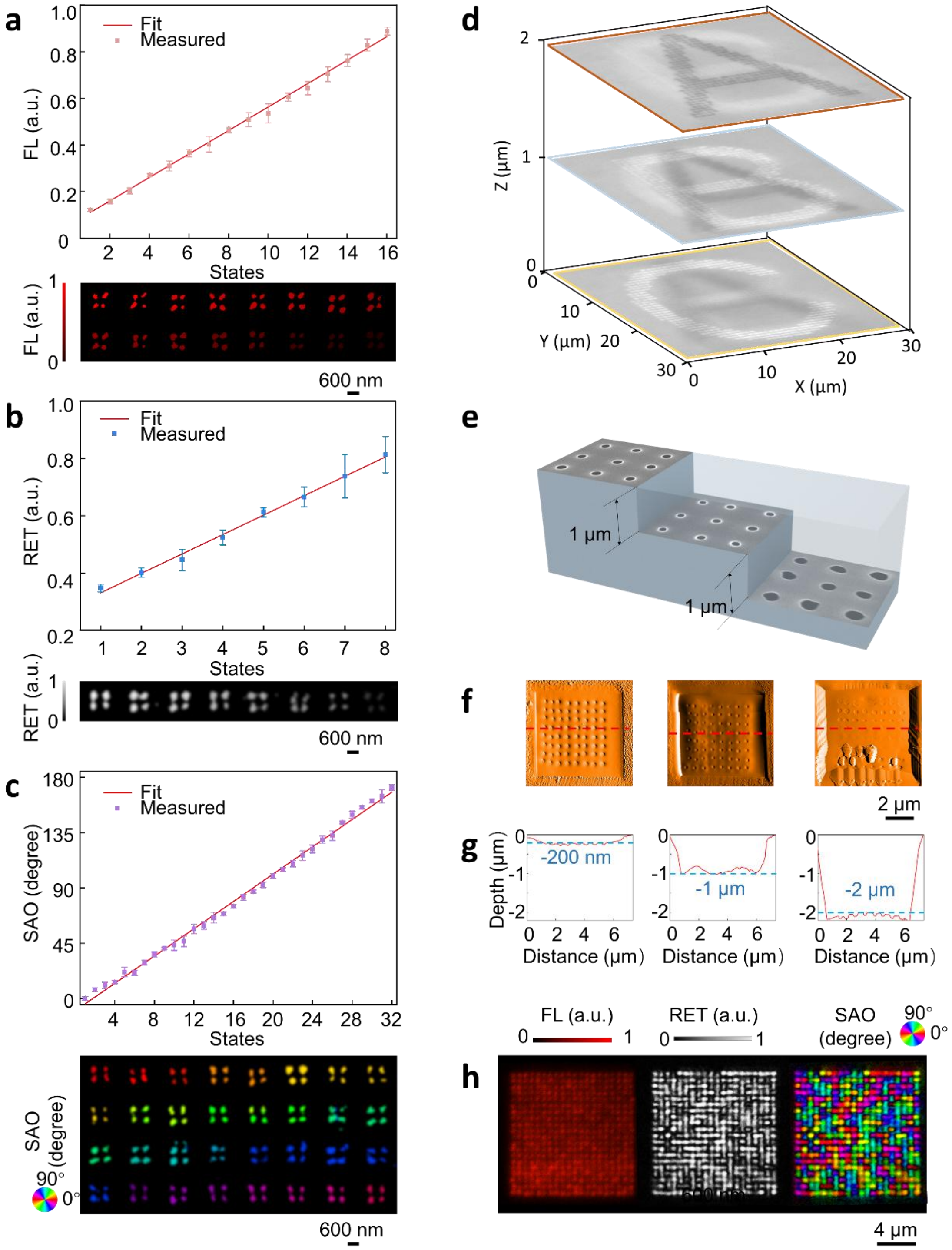

Extended Data Fig. 7 **a,b and c,** Signal distributions of FL, RET, and SAO for dot arrays with a lateral distance of 600 nm between neighboring spots. Examples of dot array profiles in the FL, RET, and SAO dimensions are shown. A top-hat algorithm is applied to these profiles. **c,inset**, Pesudo-colors indicating the SAO. **d,** Images of dot arrays staggered along the longitudinal direction with a layer separation of 1 μm. **e,** SEM images of data arrays recorded with a layer separation of 1 μm. The arrays were exposed on the surface through FIB drilling, which at the meantime destroyed the anisotropic structures and left only craters. **f,** AFM profiles corresponding to the SEM images in (**e**). **g,** AFM measurements along the dotted red lines in (**f**). **h,** FL, RET, and SAO profiles from a single dot array with a 600 nm separation between dots.

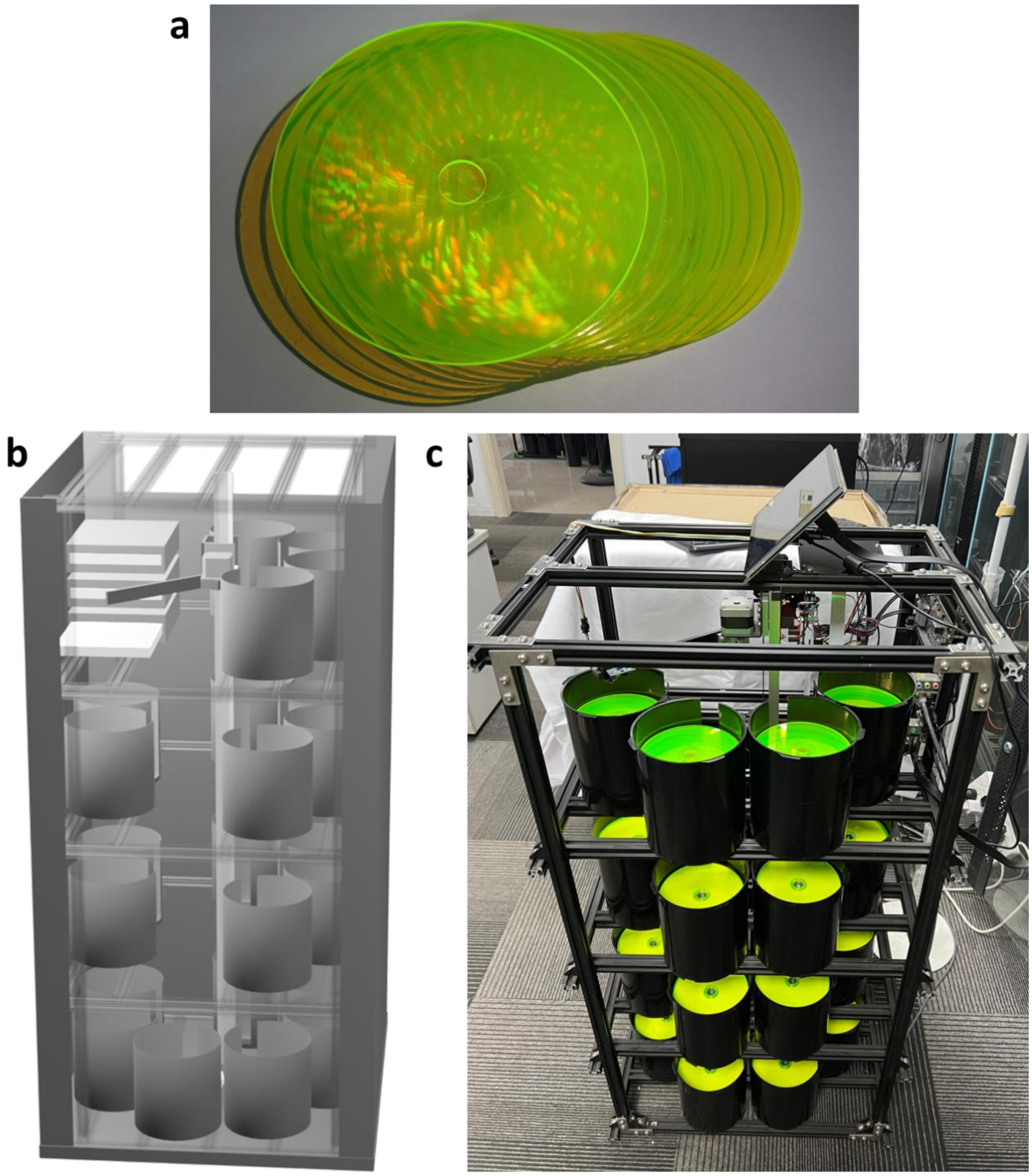

Extended Data Fig. 8 **a,** Stack of data-written 6D high-fidelity discs. **b,** Three-dimensional design of the optical disc library body. **c,** A disc library containing 6D high-fidelity discs.

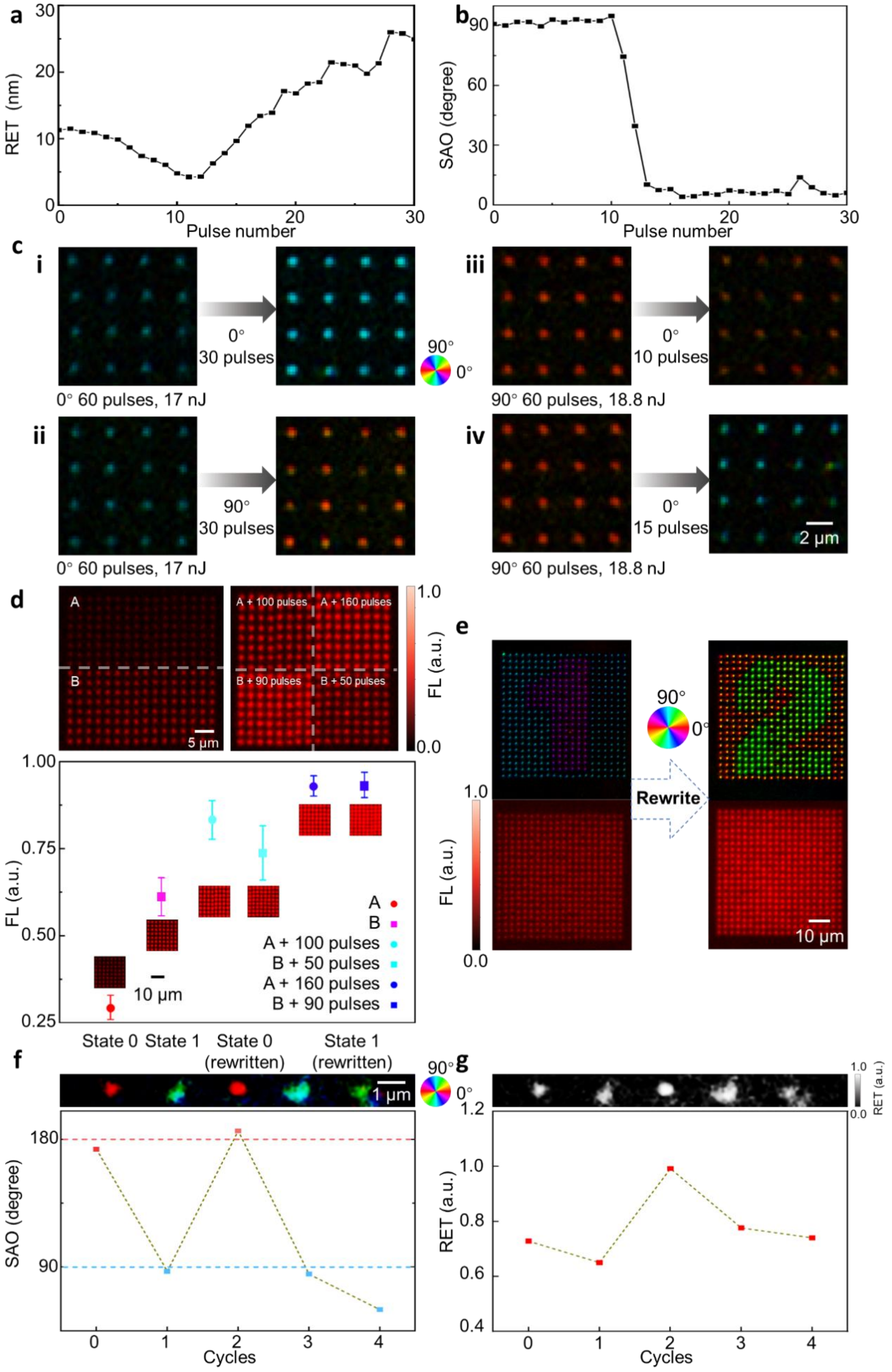

Extended Data Fig. 9 **a**, Retardance of the voxels as a function of pulse number of the rewriting beam. **b**, Slow axis orientation of the voxels as a function of pulse number of the rewriting beam. The pulse energy was 18.5 nJ. **c**, Voxel arrays with lower retardance were rewritten to higher retardance with the (**i**) same and (**ii**) different slow axis orientations. Voxel arrays with higher retardance were rewritten to lower retardance with the (**iii**) same and (**iv**) different slow axis orientations. **d**, (left) Two voxel arrays induced by different pulse energies, (right) rewriting the data voxel arrays by four different pulse sequences. (bottom) The corresponding fluorescence signals of the dot arrays. **e**, Birefringent and fluorescence images before and after rewriting. **f** and **g,** SAO and RET of a data voxel during the rewriting process as a function of cycle number. Pseudo colors in inset indicate the slow axis orientations.

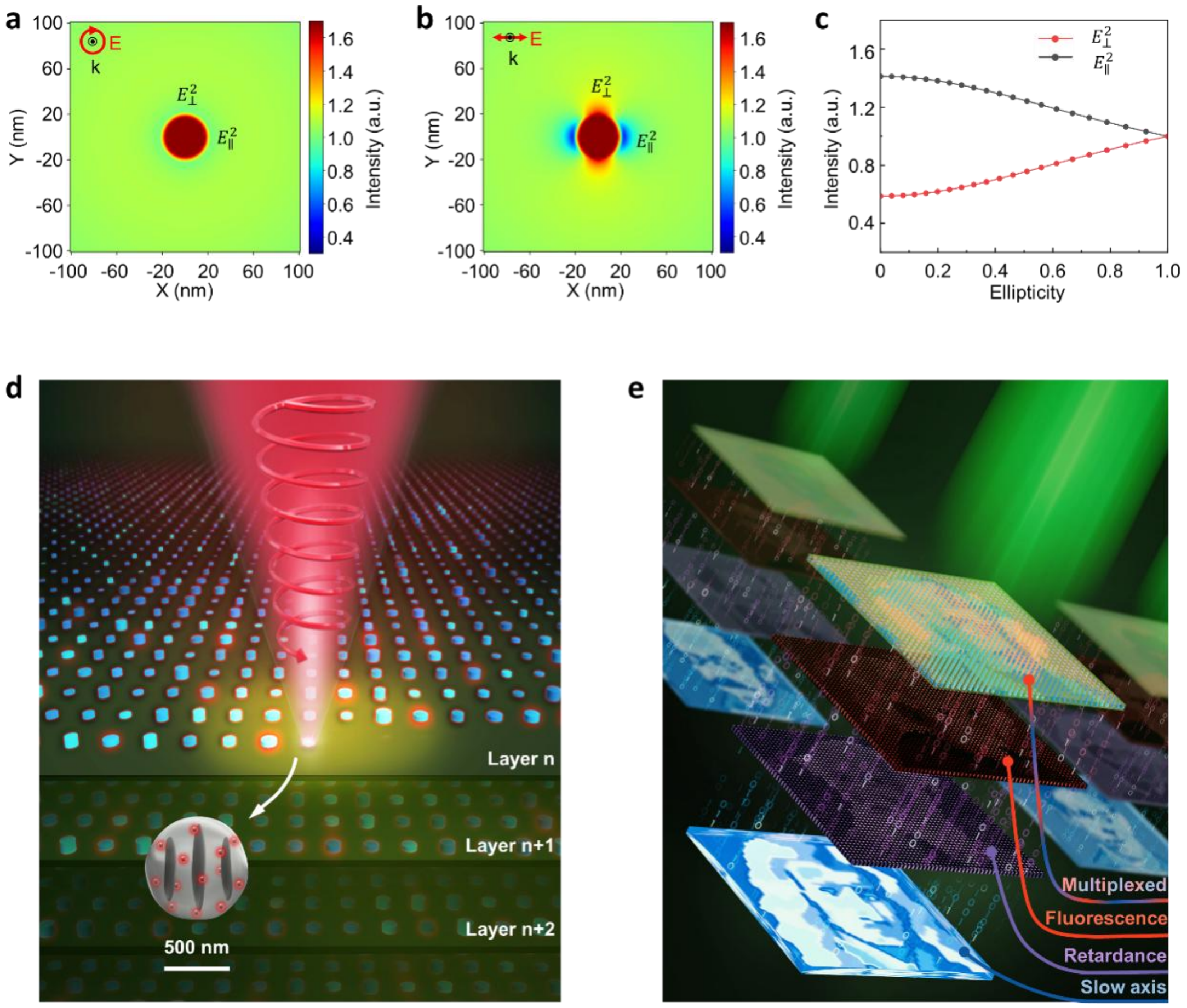

Extended Data Fig. 10 **a-c,** FDTD simulation results of light-matter interactions. **a** and **b**, Display the light intensity distributions when focused light interacts with seed nanovoids under different polarization conditions: **(a)** circularly polarized and **(b)** linearly polarized. **c**, Shows the variation in light intensity at two interfaces with changing ellipticity. Simulation parameters: wavelength: 1,030 nm; pulse duration: 220 fs; numerical aperture: 0.7; radius of seed structure: 20 nm; refractive index of seed: 1; refractive index of PAN: 1.52. Software: Lumerical FDTD Solutions. **d,** Schematic of 6D data writing. Inset presents a top-view schematic of the data voxel, featuring nanoplanes (black planes), fluorophores (dots with red glow). The glow, height, and color of each cylinder correspond to fluorescence, retardance, and slow axis orientation, respectively. **e,** Schematic of 6D multiplexing. Decrypted portraits of nine prominent scientists, encoded using fluorescence, retardance, and slow axis orientation across three multiplexed data layers.

**Supplementary information for**

# High-fidelity six-dimensional optical data storage

# Content

## S1. Visual representation

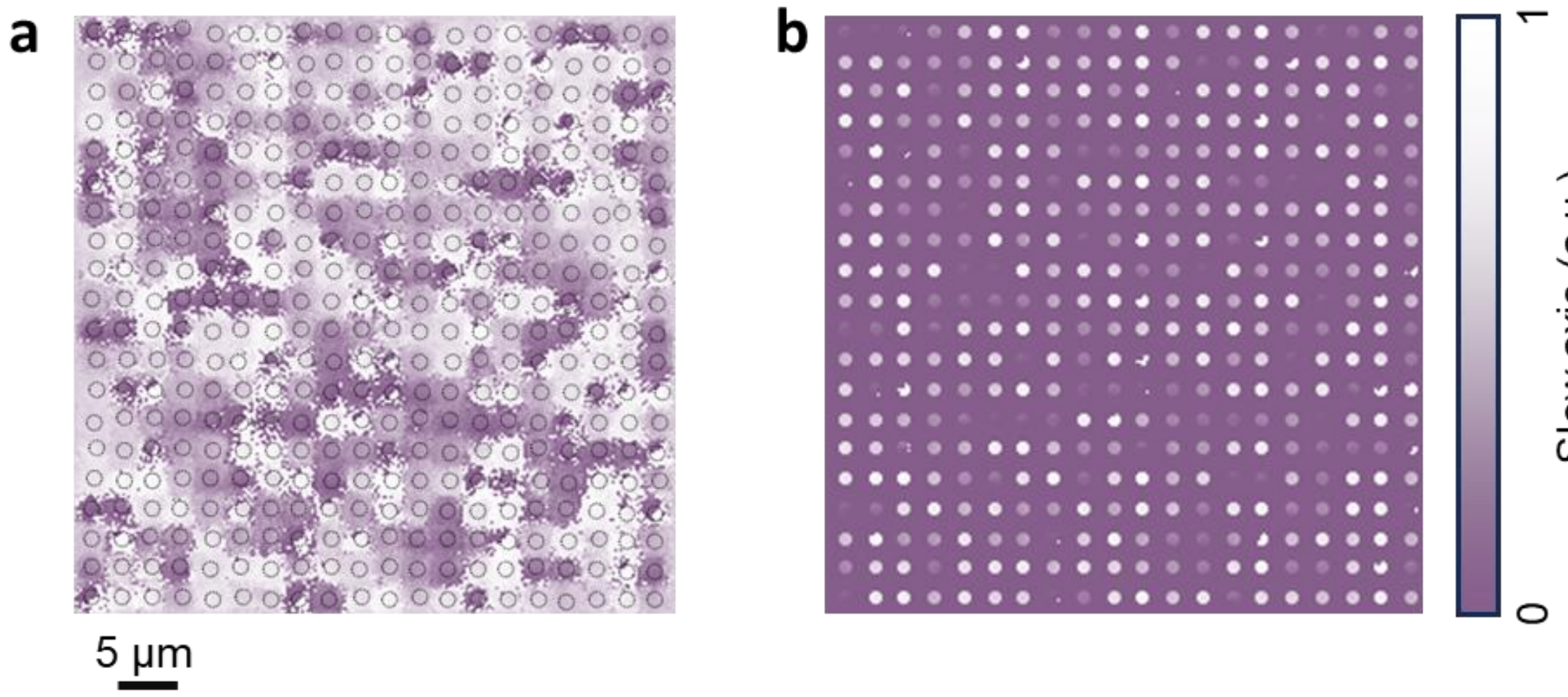

Fig. S1 **a**, Raw measurement profile of slow axis orientation. **b**, Extracted slow axis orientation profile derived from the contours of the dots, as determined by retardance measurements.

## S2. Attempt of 6D writing in fused silica

The 6D strategy was initially applied to fused silica. The fused silica samples were examined using an optical microscope to determine the modification threshold. The laser threshold disparity in fused silica is approximately 15%, due to changes in ellipticity (Fig. S2a). The signal domain created in fused silica is relatively small, and possesses a distinct tobacco pipe-like geometry (Fig. S2b). Under the same laser intensity over the threshold, the fluorescence decreases with increasing ellipticity until the ellipticity reaches approximately 0.8 (Fig. S2c). The fluorescence intensity under linear polarization was 176% higher than under circular polarization at a pulse energy of 162 nJ. These phenomena can be attributed to the lower efficiency of circularly polarized light in multi-photon absorption compared to linearly polarized light. Fewer luminescent defects were generated due to reduced energy deposition. The retardance values are significant, indicating the generation of nanogratings with form birefringence properties. In cases where $0 < \text{ellipticity} < 0.7$, both fluorescence and retardance are proportional to $I_a$ (Fig. S2d and e). Consequently, these two quantities are determined by $I_a$ simultaneously, which does not allow for the demonstration of 6D multiplexing.

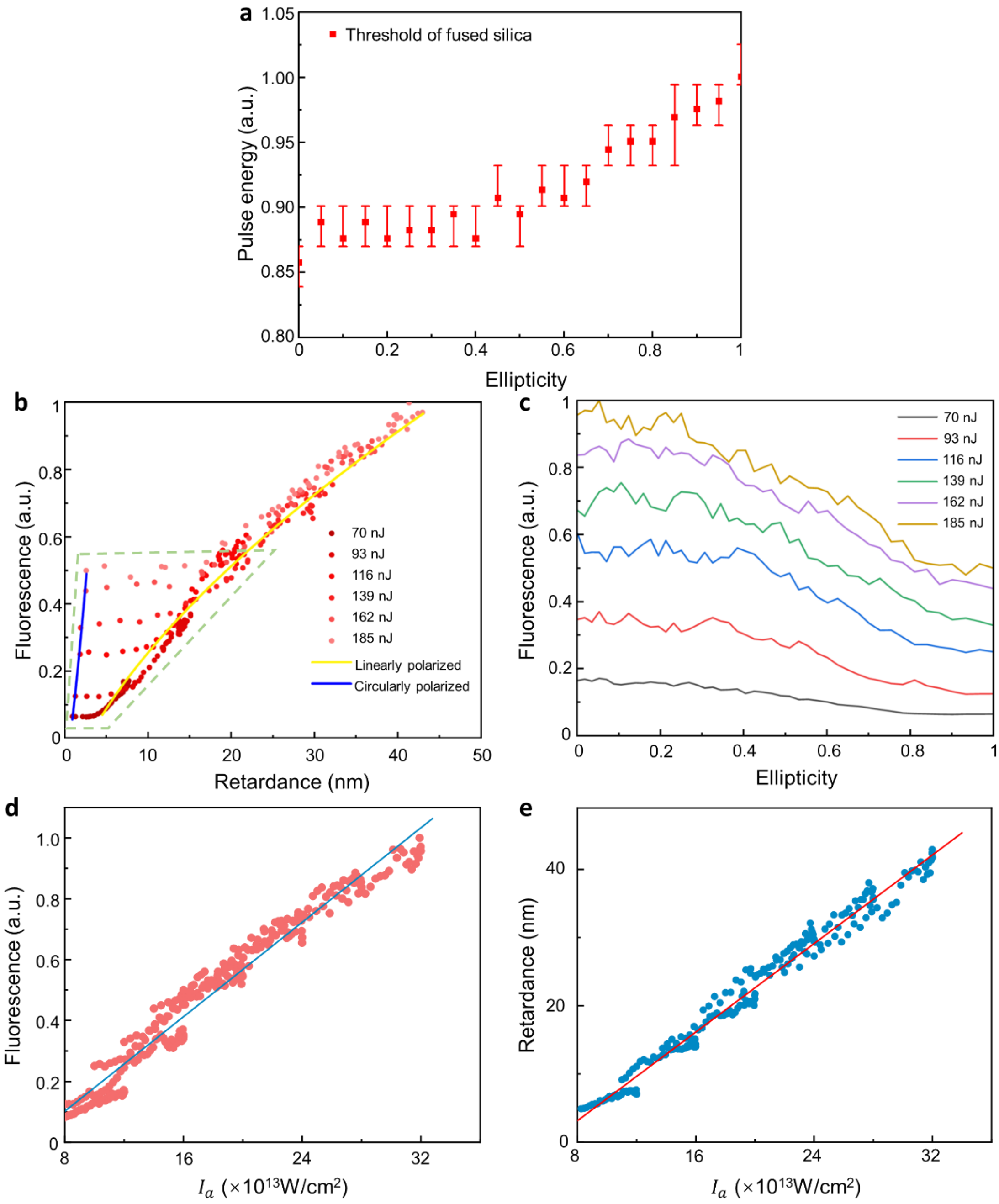

Fig. S2 Results obtained in fused silica. **a**, The single pulse modification threshold as a function of ellipticity. **b**, Distribution of fluorescence intensity and phase retardance values in a two-dimensional plane, obtained by varying ellipticity and laser intensity with a constant orientation angle at 0°. Dots of the same color represent voxels induced by the same light intensity but varying ellipticity. The ellipticity of the beam changed from 1 to 0 from left to right at the same intensity. **c**, Change of fluorescence as a function of ellipticity for varying pulse energy. **d**, Fluorescence intensity and **e**, retardance of modifications in fused silica as a function of $I_a$. 0 < ellipticity < 0.7. The number of pulses is 1,000 in b-e.

## S3 Mechanism Simulation

To investigate the polarization-dependent fluorescence response of materials under femtosecond laser irradiation, we performed numerical simulations of the temporal evolution of free-carrier density based on a rate-equation model. Under laser irradiation, the free-carrier density $N$ in the conduction band evolves according to the following rate equation[4]:

$$\frac{\partial N}{\partial t} = \widetilde{N_\mathrm{V}}\sigma_n I^n + \gamma(I) N \frac{\widetilde{N_\mathrm{V}}}{N_\mathrm{V}} \tag{1}$$

where $N$ is the free carrier density, $\widetilde{N_\mathrm{V}}$ denotes the effective density of states in the valence band, and $N_\mathrm{V}$ is the total valence electron density. The first term on the right-hand side represents multiphoton ionization (MPI), in which electrons are directly excited from the valence band to the conduction band by the simultaneous absorption of $n$ photons. The corresponding ionization rate scales as $I^n$, where $\sigma_n$ is the $n$-photon ionization cross section.

The second term accounts for avalanche ionization (AI), which describes the impact ionization process whereby energetic free carriers gain sufficient kinetic energy from the laser field to liberate additional electrons. The avalanche coefficient $\gamma(I)$ depends on the laser intensity and polarization, and is expressed as[5]:

$$\gamma(I) \approx \frac{\varepsilon_\mathrm{osc}}{J_i}\left(\frac{2\omega_0 v_\mathrm{eff}}{\omega_0^2 + v_\mathrm{eff}^2}\right) \tag{2}$$

$$\varepsilon_\mathrm{osc} = (1+\alpha^2)\frac{e^2}{4m\omega_0^2}\cdot\frac{2I}{c\varepsilon_0} \tag{3}$$

where $e$ is the elementary charge, m is the electron mass, $\omega_0$ is the laser angular frequency, $c$ is the speed of light, and $\varepsilon_0$ is the vacuum permittivity, $v_\mathrm{eff}$ is the effective collision frequency, $J_i$ is the ionization potential (the energy gap). The parameter $\alpha$ characterizes the polarization state of the laser: $\alpha = 0$ corresponds to linear polarization, $\alpha = 1$ corresponds to circular polarization, and intermediate values represent elliptical polarization. The factor $(1 + \alpha^2)$ reflects the enhancement of electron acceleration under different polarization states.

To numerically solve the carrier density evolution, the rate equation is discretized in the time domain using an explicit integration scheme. The simulation proceeds as follows:

Parameter Initialization:

Physical constants：electron charge（$e$） and mass($m$), speed of light($c$), permittivity($\varepsilon_0$)

laser parameters : wavelength $\lambda = 1030$ nm, intensity $I$ =$10^{17}$ W/m$^2$

material parameters (bandgap energy, valence electron density) are defined by different materials.

The simulation time window is divided into $N_t = 500$ steps of duration $\Delta t = 1\,fs$.

Polarization States:

Two representative polarization conditions are considered, corresponding to $\alpha = 0$ (linear) and $\alpha = 1$ (circular). For each polarization state, the avalanche coefficient $\gamma(I)$ is calculated according to the above formula.

MPI Contribution:

The contribution from multiphoton ionization is evaluated as a constant generation rate:

$$\Delta N_{\mathrm{MPI}} = \widetilde{N_{\mathrm{V}}} \sigma_n I^n \Delta t \qquad (4)$$

which is accumulated at each time step.

AI Contribution:

The avalanche ionization is calculated iteratively as

$$\Delta N_{\mathrm{AI}} = \gamma(I) N(t - \Delta t) \frac{\widetilde{N_{\mathrm{V}}}}{N_{\mathrm{V}}} \Delta t \quad (5)$$

which depends on the existing carrier density. This recursive process models the exponential growth of free carriers due to impact ionization.

Total Carrier Density:

The total density is updated at each time step as

$$N(t) = N_{\mathrm{MPI}}(t) + N_{\mathrm{AI}}(t) \quad (6)$$

$\sigma^{(K)}$ characterizes the multi-photon ionization in dense media[6]:

$$\sigma^{(K)} = \frac{2\omega}{9\pi} \left(\frac{m_e^* \omega}{\hbar}\right)^{\frac{3}{2}} D\left(\sqrt{2\left(K - \frac{E_g}{\hbar\omega}\right)}\right) I_0^{-K} \quad (7)$$

where $m_e^*$ is the reduced mass for electron-hole pair, $D(x)$ is the Dawson integral[6] , $q_e$ is the e electron charge, and[6]

$$I_0 = \frac{8 m_e^* \omega^2 E_g \varepsilon_0 c n_0}{exp(2) q_e^2} \quad (8)$$

Based on the theoretical formula, the calculated absorption cross section of fused silica is of the same order of magnitude as those reported in previous studies[7], thereby validating the accuracy of the proposed theory.

In fused silica, the relevant material parameters obtained from theoretical calculations are as follows:

$$\sigma_{\mathrm{lin}} = 7.5\times10^{-100}\ \mathrm{s}^{-1} m^{13}/\mathrm{W}^8$$

$$\sigma_{\mathrm{circ}} = 2\times10^{-100}\ \mathrm{s}^{-1} m^{13}/\mathrm{W}^8$$

The carrier densities generated in fused silica under linearly and circularly polarized irradiation, obtained from numerical simulations based on Eqs. (1)–(3), are shown in

Fig. S3 and Table S1.

Table S1. The carrier densities generated in fused silica under linearly and circularly polarized irradiation.

| | MPI($m^{-3}$) | AI($m^{-3}$) | Totol($m^{-3}$) |
|---|---|---|---|
| Linear | $1.2128\times10^{29}$ | $1.6361\times10^{26}$ | $1.2145\times10^{29}$ |
| circular | $3.2342\times10^{28}$ | $8.7321\times10^{25}$ | $3.243\times10^{28}$ |

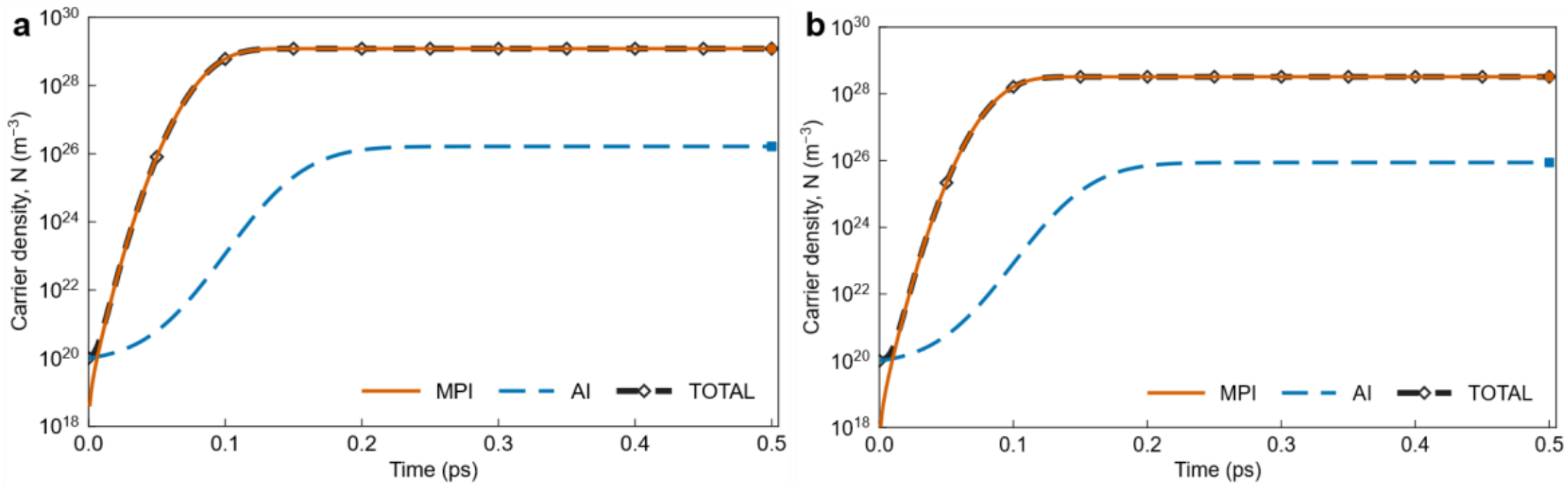

Fig. S3 Temporal evolution of the carrier densities. **a, b,** Time-dependent carrier concentrations in fused silica produced by multiphoton ionization (MPI) and avalanche ionization (AI), together with the total carrier concentration, under linear(**a**) and circularly (**b**) polarization laser irradiation. A logarithmic scale is used for the ordinate.

Following the measurement methods reported in previous studies, the transmittance of the organic material was measured. Using the fitting formula for fused silica provided therein[8]

$$T(E) = T_0(1 + (K-1)\sigma(K)E^{K-1})^{-\frac{1}{K-1}} \quad (9)$$

the absorption cross section of the organic material was subsequently determined(Fig.S4):

$$\sigma_{lin} = 9.998e^{-13}\ s^{-1}m^1/W^2$$

$$\sigma_{circ} = 7.251e^{-13}\ s^{-1}m^1/W^2$$

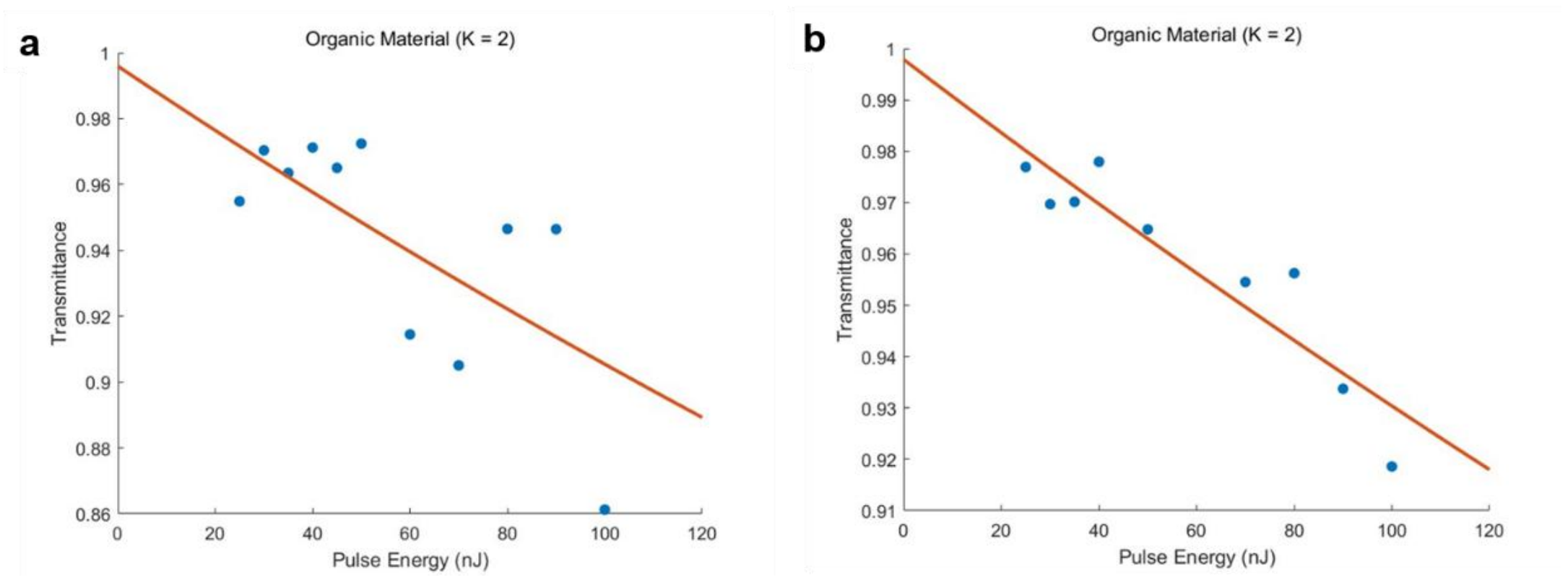

Fig.S4 Transmittance versus laser processing energy in the organic material. **a,** Experimental points and fitted curves under linearly polarized laser irradiation. **b,** Experimental points and fitted curves under circularly polarized laser irradiation.

The carrier densities generated in the organic material under linearly and circularly polarized irradiation, obtained from numerical simulations based on Eqs. (1)–(3), are shown in Fig. S5 and Table S2.

Table S2. The carrier densities generated in organic material under linearly and circularly polarized irradiation.

| | MPI($cm^{-3}$) | AI($cm^{-3}$) | Totol($cm^{-3}$) |
|---|---|---|---|
| Linear | $2.1060\times10^{13}$ | $1.0087\times10^{15}$ | $1.0298\times10^{15}$ |
| circular | $1.5271\times10^{13}$ | $1.0175\times10^{15}$ | $1.0328\times10^{15}$ |

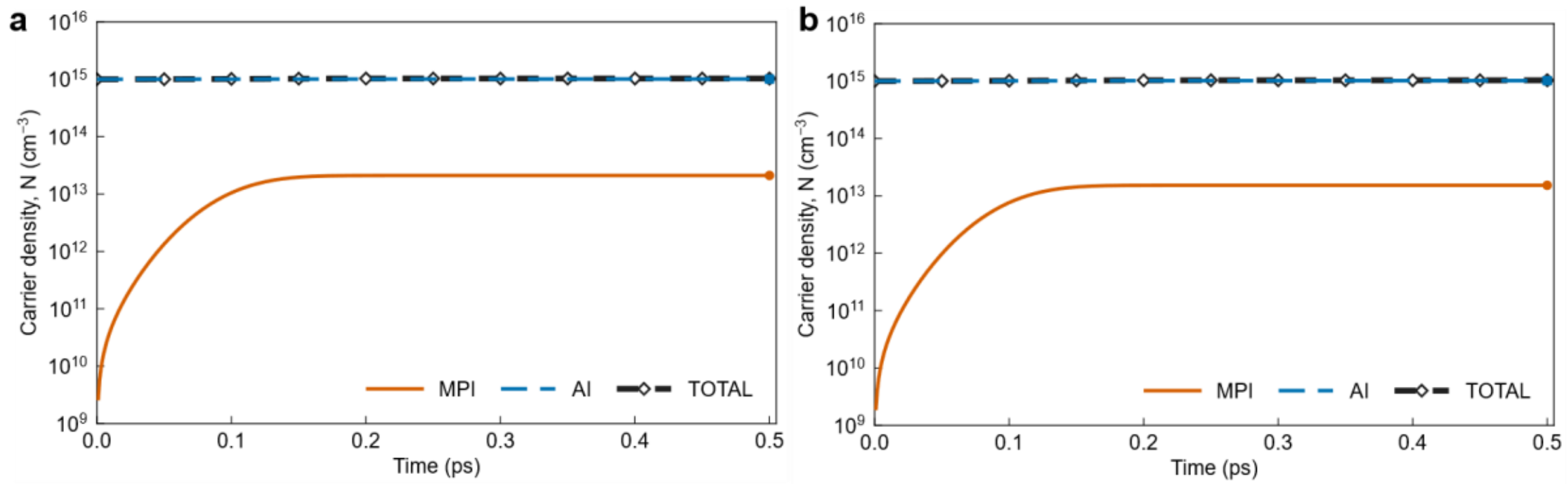

Fig.S5 Temporal evolution of the carrier densities. **a, b,** Time-dependent carrier concentrations in organic material produced by multiphoton ionization (MPI) and avalanche ionization (AI), together with the total carrier concentration, under linear(**a**) and circularly (**b**) polarization laser irradiation. A logarithmic scale is used for the ordinate.

We solved the rate equation using an explicit time-integration scheme. Two material systems were investigated, wide-bandgap fused silica and $MAPbBr_3$-doped polymers.

The multiphoton absorption order $K$ and the corresponding cross-section $\sigma^{(K)}$ for silica were calculated following the Dawson-integral formalism[3,4], yielding values consistent with previously published results[4], thus validating our model. For fused silica, the calculated effective absorption cross-sections for linear and circular polarization are $7.5 \times 10^{-100}$and $2 \times 10^{-100}\mathrm{s}^{-1}\mathrm{m}^{13}\mathrm{W}^{8}$, respectively. For the organic material, the two-photon absorption cross-sections were independently obtained by fitting the transmittance data[5], yielding $9.998 \times 10^{-13}$ and $7.251 \times 10^{-13}\ \mathrm{s}^{-1}\mathrm{m}^{1}\mathrm{W}^{2}$ for linear and circular polarization, respectively.

The simulation results (Fig.S3-5 and Table S1-2) show that, in fused silica, the absorption cross-section for linear polarization is approximately 3.75 times that for circular polarization, leading to a substantially higher total carrier density under linear excitation $(1.21 \times 10^{29}\mathrm{m}^{-3})$ than under circular excitation $(3.24 \times 10^{28}\mathrm{m}^{-3})$. This pronounced difference directly explains the strong polarization-dependent fluorescence observed in fused silica (Fig. S2). However, for the $MAPbBr_3$ system, the total carrier densities under linear and circular polarizations are nearly identical (both around $10^{15}\mathrm{cm}^{-3}$ ), consistent with the negligible fluorescence difference observed experimentally. Overall, the simulation results indicate that the influence of polarization on carrier excitation is governed by the photon absorption order, which depends on both the material bandgap and the excitation wavelength. This provides the physical basis for decoupling fluorescence from the polarization state by reducing the photon absorption order.

## S4. Material selection process

We selected carbon-containing polymers, such as polyacrylonitrile (PAN), polyvinyl butyral (PVB), and polyvinylpyrrolidone (PVP), as potential substrate materials. In these substrates, fluorescent carbon dots are expected to form following laser modification. After doping with $MAPbBr_3$ (0.03 M), these materials exhibited birefringence signals after laser modification (Fig. S6e-g). These observations demonstrate the applicability of our technique.

Although all three polymers can be measured under microscopy, PVB substrate exhibited opaque brown (Fig. S6b), while PVP displayed softness and high viscosity (Fig. S6c). Consequently, we selected PAN as our substrate. While the PAN substrate doped with $FAPbBr_3$ (0.03 M) exhibited birefringence signals (Fig. S6h), its opaque brown appearance made it difficult to increase the thickness (Fig. S6d).

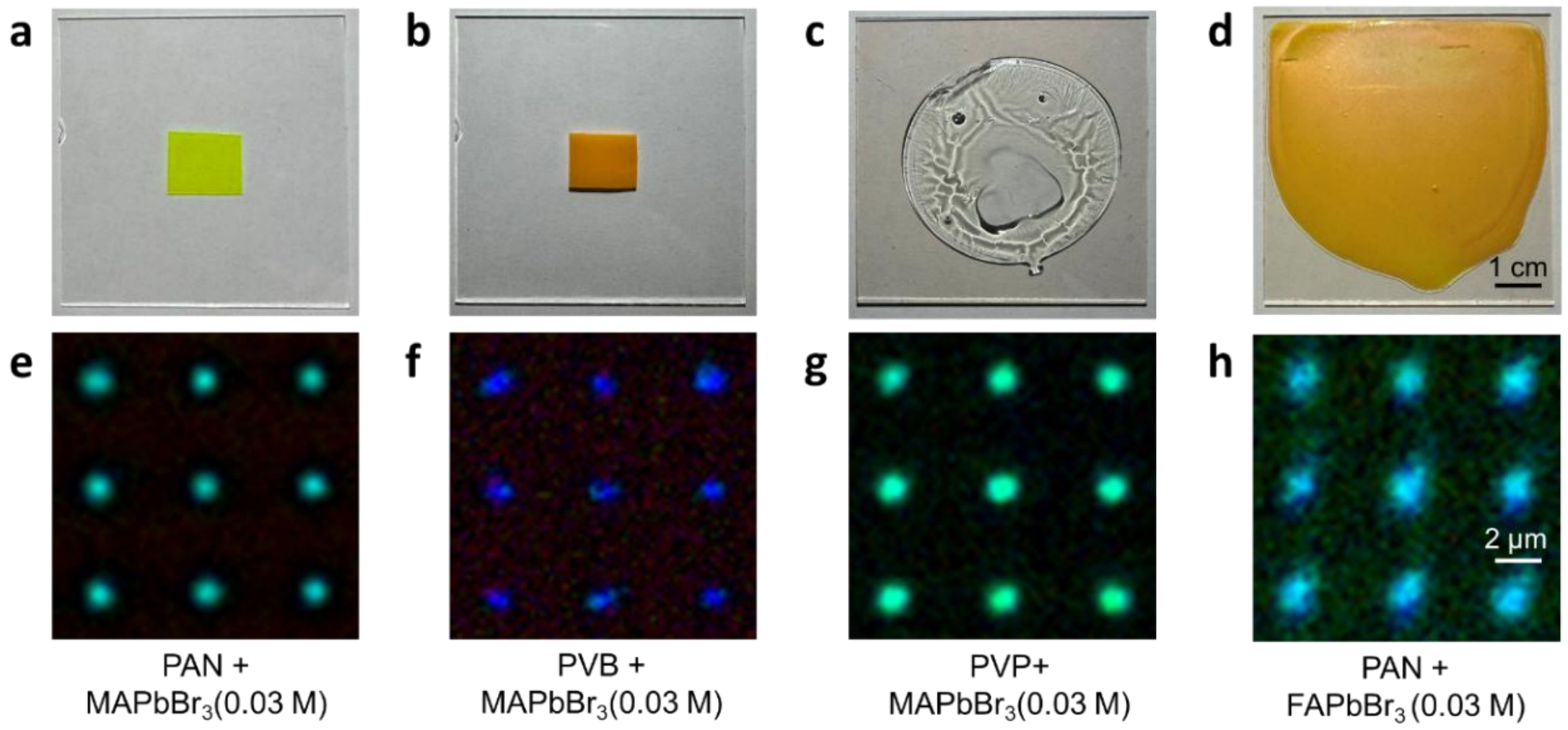

Fig. S6 **a-d,** Optical images of samples and **e-h,** corresponding birefringence measurements of laser-modified voxels, with varied polymers and dopants.

PAN substrates were doped with varying concentrations of $MAPbBr_3$ for examination (Fig. S7). For pristine PAN and PAN doped with 0.01 M $MAPbBr_3$ (Fig. S7a, b), the generated voxels exhibited birefringence profiles typical of a void structure (Fig. S7e, f, i, and j). However, when PAN was doped with 0.1 M $MAPbBr_3$, the material became non-transparent (Fig. S7d), rendering microscope measurements unfeasible (Fig. S7h). Structures generated in PAN with a concentration of 0.03 M $MAPbBr_3$ displayed uniform and smooth profiles for both slow axis orientation and retardance (Fig. S7g, k). Consequently, PAN doped with 0.03 M $MAPbBr_3$ was selected as our candidate for 6D data writing.

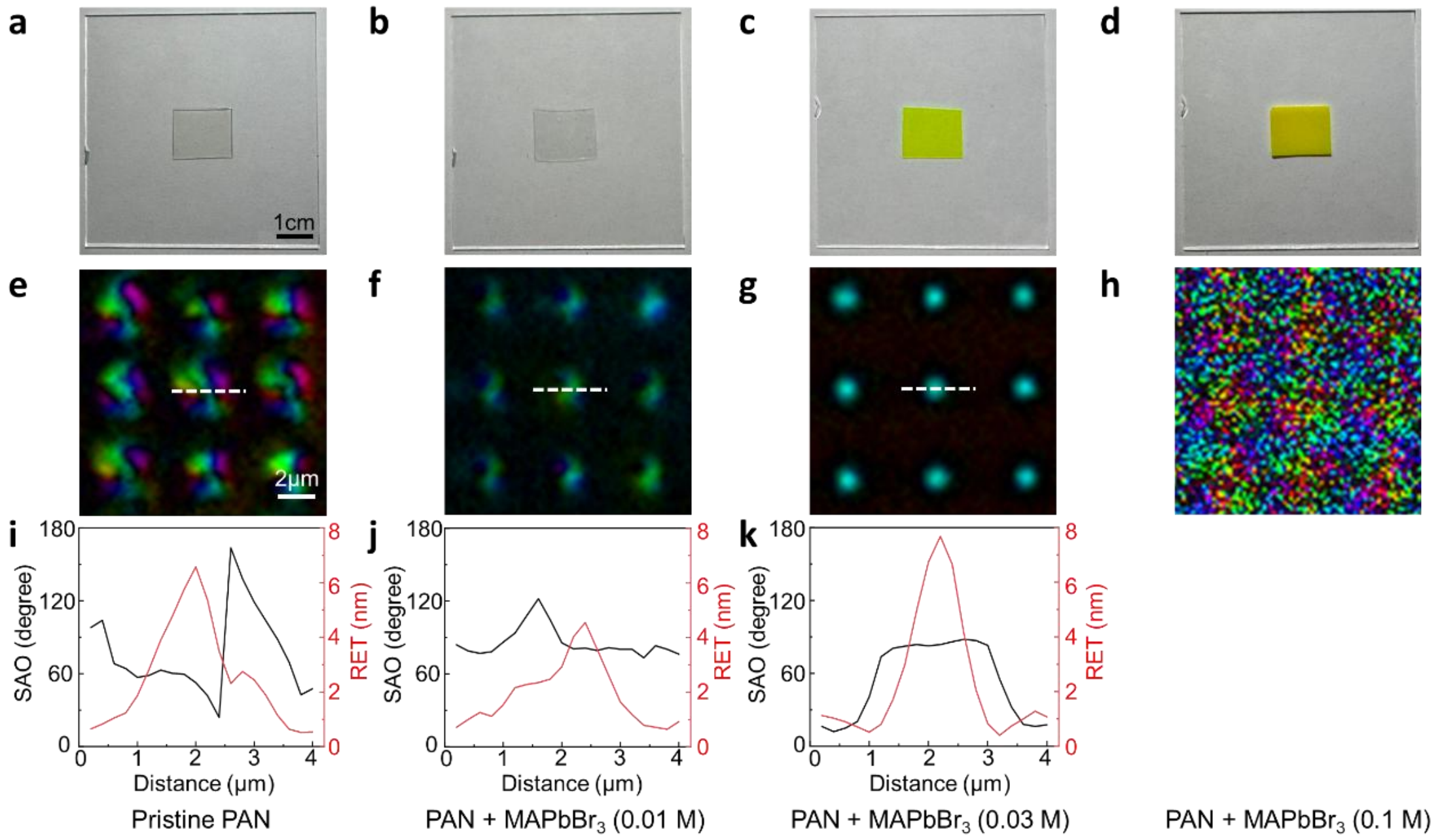

Fig. S7 **a-d,** Optical images of samples. **e-h,** Corresponding birefringence measurements of laser-modified voxels. **i-k,** Slow axis orientation and retardance profiles derived from the white dashed lines in **e-g**.

Formamidinium bromide (FABr, 99.9%) and lead bromide ($PbBr_2$, 99.9%) were purchased from Advanced Election Technology Co., Ltd. Polyvinyl butyral (PVB, average Mw 80,000) was obtained from Aladin. Polyvinylpyrrolidone (PVP, average

Mw 58,000) was obtained from Alfa Aesar. For the formamidinium lead bromide ($FAPbBr_3$) solution, the precursor concentration is 0.1 M. For the PVB sample, the PVB concentration is 250 mg/mL. For the PVP sample, the PVP concentration is 600 mg/mL. The other preparation procedures were identical to those used for the fabrication of $MAPbBr_3$-doped PAN.

## S5. Applicability of the 6D multiplexing

Different types of modifications can be generated in glass by changing laser parameters, including refractive index change (type I), nanogratings (type II), voids (type III)[1] and elliptical nanovoids (type X)[2]. These modifications exhibit distinct fluorescent properties due to the generation of different defects[3]. In our experiments with silica, the fluorescence of non-bridging oxygen hole centers (NBOHC) exhibited the highest intensity compared to other defects. It is speculated that the circularly polarized light, which is less efficiently absorbed, primarily contributes to the type I modification, which is the weakest modification induced by femtosecond laser machining in glass. And the type I modification corresponds to the generation of NBOHC[3]. In other words, circularly polarized light is less absorbed but more efficient in generating NBOHC. As a consequence, the decline in fluorescence intensity slows down as the ellipticity approaches one (Fig. S2c).

Instead of the one-dimensional subspace generated by a linearly polarized beam in the fluorescence-retardance space (represented by the yellow curve in Fig. S2b), the application of the light modulation approach to fused silica creates a two-dimensional signal domain (indicated by the green dotted trapezoid in Fig. S2b). Despite the small signal domain created, it holds immense potential. It is anticipated that by adding low-bandgap materials, such as metal photosensitive ions, glasses with high robustness can turn into an eternal medium for 6D data storage.

Besides glass as a promising candidate, polymers also demonstrate significant potential. Our preliminary results have experimentally shown that different substrates and dopants can also facilitate birefringence modification (Fig. S7, Fig. S8). Considering the inherent properties of polymers, fluorescence carbon dots can be induced through femtosecond laser irradiation. And a variety of dopant candidates facilitates two-photon excitation. We anticipate that our 6D multiplexing approach will be applicable to a broad spectrum of polymers, enabling high-dimensional data multiplexing in the near future.

| | Pristine | $MAPbBr_3$ doped | $FAPbBr_3$ doped |
|---|---|---|---|
| PAN | ✗ Pulse energy(0-30nJ) | ✓ | ✓ |
| PVB | ✗ Pulse energy(0-45nJ) | ✓ | \ |
| PVP | ✗ Pulse energy(0-16.5nJ) | ✓ | \ |

✗ Not applicable to six-dimensional storage

✓ Applicable to six-dimensional storage

\ Didn't investigate

Fig. S8 A summary of polymer and dopant candidates.